\documentclass[aps,pra,amsmath,amsfonts,twocolumn,10pt,letterpaper,nofootinbib,floatfix]{revtex4-2}

\usepackage{graphicx} 
\usepackage{hyperref}
\usepackage{amsthm}
\usepackage{mathtools}
\usepackage{relsize}
\usepackage{lipsum}
\usepackage{xcolor}
\usepackage{algorithm, algpseudocode}
\usepackage{makecell}
\usepackage{enumerate}

\newcommand{\F}{\mathbb{F}_2}
\newcommand{\Z}{\mathbb{Z}_4}
\newcommand{\inner}[2]{\left\langle #1, #2 \right\rangle}

\DeclareMathOperator{\majo}{\mu}
\DeclareMathOperator{\majoBar}{\bar{\majo}}
\DeclareMathOperator{\Majo}{\mathcal{M}}

\DeclareMathOperator{\pauli}{\pi}

\DeclareMathOperator{\cliff}{\Gamma}
\DeclareMathOperator{\Cliff}{\mathcal{C}}

\DeclareMathOperator{\Stab}{\mathcal{S}}
\DeclareMathOperator{\Logic}{\mathcal{L}}

\DeclareMathOperator{\tr}{tr}

\DeclareMathOperator{\Frame}{\mathcal{F}}
\DeclareMathOperator{\ensemble}{\mathcal{E}}
\DeclareMathOperator*{\Exp}{\mathlarger{\mathbb{E}}}

\DeclareMathOperator{\Hilbert}{\mathcal{H}}

\DeclareMathOperator{\GL}{GL}
\DeclareMathOperator{\Sp}{Sp}
\DeclareMathOperator{\Ortho}{O}
\DeclareMathOperator{\Unit}{U}

\DeclareMathOperator{\vspan}{span}

\newcommand{\bra}[1]{\langle #1 \rvert}
\newcommand{\ket}[1]{\lvert #1 \rangle}
\newcommand{\braket}[1]{\langle #1 \rangle}

\DeclareMathOperator{\symp}{\omega}
\DeclareMathOperator{\sympU}{\symp_U}
\DeclareMathOperator{\sympL}{\symp_L}
\DeclareMathOperator{\parity}{p}
\DeclareMathOperator{\weight}{w}

\DeclareMathOperator{\braid}{BRAID}

\DeclareMathOperator{\symppauli}{\eta}
\DeclareMathOperator{\symppauliL}{\symppauli_L}

\newtheorem{lemma}{Lemma}
\newtheorem{theorem}{Theorem}
\newtheorem{proposition}{Proposition}

\newtheorem{corollary}{Corollary}
\theoremstyle{definition}
\newtheorem{definition}{Definition}

\algnewcommand\Break{\textbf{break}}

\begin{abstract}
    In quantum information science, Clifford operators and stabilizer codes play a central role for systems of qubits (or qudits). In this paper, we study their analogues for systems composed of Majorana fermions. In this case, a crucial role is played by fermion parity symmetry, which is an unbreakable symmetry present in any system with fundamentally fermionic degrees of freedom. We prove that the subgroup of parity-preserving Majorana Cliffords can be represented by the orthogonal group over the binary field $\F$, and we show how it can be generated by braiding operators and used to construct any (even-parity) Majorana stabilizer code. We also analyze the frame potential for this so-called p-Clifford group when acting on a fixed-parity sector of the Hilbert space, proving that it is equivalent to the frame potential of the ordinary Clifford group acting on the same sector.
\end{abstract}

\begin{document}

\title{The Structure of the Majorana Clifford Group}
\author{Val\'erie Bettaque}
\email{vbettaque@brandeis.edu}
\affiliation{Martin A. Fisher School of Physics, Brandeis University, Waltham, MA 02453, USA}
\author{Brian Swingle}
\email{bswingle@brandeis.edu}
\affiliation{Martin A. Fisher School of Physics, Brandeis University, Waltham, MA 02453, USA}

\maketitle

\section{Introduction}
\label{sec:introduction}

The standard model of quantum computation is usually expressed in terms of bosonic (i.e.\ commuting) degrees of freedom, but one can also formulate an essentially equivalent model in terms of fermionic (i.e.\ anti-commuting) degrees~\cite{bravyi_fermionic_quantum_computation_2002}. In the bosonic formulation, an important role is played by the Clifford group, which is a special subgroup of unitaries whose elements map between tensor products of Pauli operators~\cite{gottesman_stabilizer_codes_quantum_1997}. Here we study the fermionic analog of this structure, the \emph{Majorana Clifford group}, whose (unitary) elements act in an equally simple way on operator products of Majorana fermion modes. And while any Majorana Clifford turns out be just a Jordan-Wigner transformation away from becoming a Pauli Clifford, the non-locality of said transformation together with \emph{fermion parity superselection} places new constraints on which subset of operations can be considered to be physical.

One motivation for an analysis of this fermionic Clifford group is provided by a promising class of fault-tolerant quantum computer architectures that employs the topological braiding of Majorana fermion modes to encode logical information \cite{nayak_non_abelian_anyons_2008, tran_optimizing_clifford_gate_2020, karzig_scalable_designs_quasiparticle_2017, lutchyn_majorana_fermions_topological_2010, alicea_new_directions_pursuit_2012, aghaee_inas_al_hybrid_2023}.
While their topological structure provides these systems with some inherent degree of protection against outside noise, the concatenation of additional error-correcting (stabilizer) codes generated by Clifford operations acting on these Majorana modes will likely still be necessary to achieve adequate macroscopic resilience. \emph{Majorana stabilizer codes}, which were initially proposed by Bravyi et al.\ \cite{bravyi_majorana_fermion_codes_2010}, have since garnered a rich literature \cite{litinski_quantum_computing_majorana_2018, vijay_majorana_fermion_surface_2015, vijay_quantum_error_correction_2017, mclauchlan_fermion_parity_based_2022, mclauchlan_new_twist_majorana_2024, mudassar_encoding_majorana_codes_2024}. Such codes have advantages compared to their Pauli- and Dirac-based counterparts, most notably their natural protection against errors affecting multiple Majoranas at once, assuming they are sufficiently spatially separated. This is thanks to the aforementioned fermionic superselection rules and the resulting non-locality of observables and parity-preserving operations \cite{kitaev_unpaired_majorana_fermions_2001}. The major source of errors for Majorana codes is therefore characterized by single-mode errors, which are referred to as \emph{quasiparticle poisoning} \cite{vijay_quantum_error_correction_2017, mudassar_encoding_majorana_codes_2024}. Constructing good codes that can detect (and correct) such errors is the subject of ongoing research, to which a systematic analysis of all embedding Majorana Cliffords could prove to be a valuable contribution.

Another motivation arises from the \emph{Sachdev-Ye-Kitaev (SYK) model}, which has recently become one of the most-studied physical systems in the fields of condensed matter and quantum gravity \cite{sachdev_gapless_spin_fluid_1993, kitaev_simple_model_quantum_2015, polchinski_spectrum_sachdev_ye_2016, maldacena_remarks_sachdev_ye_2016, haldar_variational_wavefunctions_sachdev_2021}. Describing a large number of randomly coupled Majorana fermions, it manages to provide a simple toy model for holography, connecting concepts in quantum information and high energy physics. Recently we proposed a potential tensor network ansatz for the SYK model, and analyzed a qudit-based toy example of it by employing random Clifford gates \cite{bettaque_nora_tensor_network_2024}. The desire to generalize this toy model to fermionic degrees of freedom was the starting point of this paper, although we have deferred any analysis of the corresponding fermionic tensor networks to future work.

Given these motivations and the broadly important role played by the ordinary Clifford group in quantum information and physics, this paper presents a thorough analysis of its fermionic counterpart. We are hence primarily concerned with the description and manipulation of systems composed of $2n$ Majorana fermion modes $\chi_i$ $(i=1, \ldots, 2n)$, obeying
\begin{equation}
\label{eq:majorana_def}
    \chi_i^{\dagger} = \chi_i, \quad \{ \chi_i, \chi_j \} = 2 \, \delta_{ij},
\end{equation}
and equipped with a fundamental representation on a $2^n$-dimensional Hilbert space. Such systems exhibit a parity symmetry, which is generated by the Hermitian operator
\begin{equation}
    (-1)^F \propto \prod_{i=1}^{2n} \chi_i.
\end{equation}
This symmetry imposes constraints on valid physical quantities and operations, which are usually referred to as parity superselection. In general, such superselection rules dictate that for an observable $O$ (including the Hamiltonian) to be physical, it must obey
\begin{equation}
    \braket{\psi_1 | O | \psi_2} = 0,
\end{equation}
for any two states $\ket{\psi_1}$, $\ket{\psi_2}$ from different superselection sectors (here corresponding to sectors with even and odd parity). This implies that states from different sectors can't be put into coherent superpositions, and that the superselection-defining observable (in this case $(-1)^F$) must be conserved. Hence we primarily focus on \emph{p-Cliffords}, the subgroup of Majorana Cliffords whose elements commute with $(-1)^F$.

Our findings show that the group of p-Cliffords admits a projective representation in terms of binary orthogonal matrices, a subgroup of the symplectic group describing all Cliffords. Each orthogonal matrix can be generated by a small set of Householder reflections, which in the Hilbert space representation corresponds to the p-Cliffords being generated by so-called \emph{Majorana braiding operators}. Using a canonical basis for even-parity stabilizers, we show that by, adding ancilla modes, any other even stabilizer basis (and its corresponding logical operators) can be generated by applying an appropriate p-Clifford. Last but not least, we prove that the p-Clifford group can be considered to be a unitary design on a fixed-parity sector of the Hilbert space, but like the ordinary Clifford group only manages to be a 3-design.

Overall, the rest of this paper is structured as follows: In Section \ref{sec:majorana_strings} we formally define Majorana strings, prove some basic properties, and discuss connections between operator weight, operator parity and the imposed superselection rules. We also highlight the relation to Pauli strings and the resulting Hilbert space representation. Next, in Section \ref{sec:cliffords} we formally define the general (Majorana) Clifford group and identify the p-Cliffords as a subgroup, characterized in terms of orthogonal matrices over the field $\F$ and efficiently generated by braiding operators. In Section \ref{sec:stabilizers} we discuss fermionic stabilizer states and how -- with the help of an ancilla system -- they can be embedded using p-Cliffords. Fourth, in Section \ref{sec:design} we discuss the problem of randomly sampling from the p-Clifford group, analyze its frame potential, and show that the p-Clifford group is a 3-design within sectors of fixed fermion parity. We conclude with an outlook and discussion of open problems in Section \ref{sec:conclusions}.

Example implementations of the algorithms explored in this work are available as a Julia Pluto.jl notebook. The notebook can be found in the supplemental materials, or can be provided on direct request.

\textbf{Note:} At the time of writing, the authors became aware of independent research being conducted on the same topic~\cite{mudassar_encoding_majorana_codes_2024}. Their results inspired some of the terminology in Section \ref{sec:generating_set}, and a variation of their stabilizer generation algorithm using our established framework in section \ref{sec:stabs_from_cliffords}.

\section{Majorana Strings}
\label{sec:majorana_strings}

To describe operator products of fundamental Majorana fermion modes and the superoperators that map between them, it is most convenient to use the language of vector spaces over the binary field $\F \equiv (\{0, 1\}, +, \cdot)$. This construction follows analogous (and equivalent) programs applied to ordinary Pauli strings \cite{gross_hudsons_theorem_finite_2006}.

\subsection{General Properties}

A reasonable definition for Hermitian strings made of some even number of elementary Majorana mode operators is as follows:

\begin{definition}[Hermitian Majorana String]
    Given $2n$ independent Majorana fermion operators $\chi_i$ satisfying \eqref{eq:majorana_def}, a general Hermitian \emph{Majorana string} $\majo(v)$ is defined (up to an overall sign) to be the normal-ordered product of said operators:
    \begin{equation}
    \label{eq:majo_string}
        \majo(v) \coloneqq (i)^{v^T \sympL v} \cdot \chi_1^{v_1} \chi_2^{v_2} \cdots \chi_{2n}^{v_{2n}}, \quad v \in \F^{2n},
    \end{equation}
where we implicitly defined the lower-triangular matrix
\begin{equation}
\label{eq:symp_matrix_lower}
    \omega_L \coloneqq \begin{pmatrix}
        0 & 0 & \cdots & 0 & 0 \\
        1 & 0 & \cdots & 0 & 0 \\
        \vdots & \vdots & \ddots & \vdots & \vdots \\
        1 & 1 & \cdots & 0 & 0 \\
        1 & 1 & \cdots & 1 & 0
    \end{pmatrix} \in \F^{2n \times 2n}.
\end{equation}
For the sake of completeness we also identify its upper-triangular counterpart as $\sympU \equiv \sympL^T$.
\end{definition}

It is straightforward to show that this definition indeed satisfies $\majo(v)^\dagger = \majo(v)$ by reordering the operators after taking the adjoint. All expressions involving elements of the finite field $\F$ are understood to be evaluated mod 2. This also implies that complex phases with elements of $\F$ in the exponent don't obey the usual algebraic rules\footnote{The reason for this is the incompatibility between $\F$ having addition mod 2 and $(i)^a = (i)^{a+4}$ implying addition mod 4. The same issue also plagues the definition of qubit Pauli strings.} and should therefore be treated with care.

With that we can define the equivalent of the Pauli group as the span of all (not necessarily Hermitian) Majorana strings:

\begin{definition}[Majorana Group]
    The \emph{Majorana group} $\Majo_{2n}$ is the set of all Majorana strings of length $2n$:
    \begin{equation}
    \label{eq:majorana_group}
        \Majo_{2n} \coloneqq \{ (i)^a \cdot \majo(v) \, | \, a \in \Z, v \in \F^{2n} \}.
    \end{equation}
    $\Majo_{2n}$ is a non-abelian group under operator multiplication, and an irreducible projective representation of the abelian group $\Z \times \F^{2n}$ (see Lemma \ref{thm:majo_irreducible}).
\end{definition}

The group structure of $\Majo_{2n}$ follows directly from the composition rule for two (Hermitian) Majorana strings:
\begin{equation}
\label{eq:majorana_composition}
    \majo(v) \majo(v') = \zeta(v, v') \cdot \majo(v + v').
\end{equation}
Here $\zeta(v, v')$ is some non-trivial phase that arises partly due to the aforementioned algebraic quirks of the definition, and which can be shown to be of the form
\begin{equation}
\label{eq:composition_coeff_1}
    \zeta(v, v') \equiv (-1)^{v^T \sympL v' + f(v, v')} \, (i)^{v^T \omega v'},
\end{equation}
where we also identified the symmetric function
\begin{align}
\begin{split}
\label{eq:composition_coeff_2}
    f(v, v') \equiv{}& (v^T \omega_L v) (v'^T \omega_L v') \\
    &+ v^T \omega v' (v^T \omega_L v + v'^T \omega_L v' + 1).
\end{split}
\end{align}

Both expressions $\zeta(v,v')$ and $f(v,v')$ involve a new symmetric\footnote{Analogous matrices over fields of higher order (corresponding to qudits with $d > 2$ prime) are actually skew-symmetric, but for $\F$ this makes no difference.} matrix $\symp$ which arises in the derivation as the sum of $\sympL$ and its transpose $\sympU$:
\begin{equation}
\label{eq:symp_matrix_def}
    \symp \coloneqq \symp_L + \symp_U = \begin{pmatrix}
        0 & 1 & \cdots & 1 & 1 \\
        1 & 0 & \cdots & 1 & 1 \\
        \vdots & \vdots & \ddots & \vdots & \vdots \\
        1 & 1 & \cdots & 0 & 1 \\
        1 & 1 & \cdots & 1 & 0
    \end{pmatrix} \equiv I^c.
\end{equation}
Here $I^c$ denotes the \emph{matrix complement} of the identity.

\begin{definition}[Complement]
For any $n \times m$ matrix $A$ over $\F$, we define its \emph{matrix complement} $A^c$ as the matrix one gets when exchanging $0 \leftrightarrow 1$. This is equivalent to writing
    \begin{equation}
        A^c \equiv J - A,
    \end{equation}
    where $J$ is the $n \times m$ matrix with all its entries equal to 1. \emph{Vector complements} are defined analogously using the corresponding all-ones vector:
    \begin{equation}
    \label{eq:all_ones_vector}
        j \coloneqq (1, 1, \ldots, 1)^T.
    \end{equation}
\end{definition}

While \eqref{eq:composition_coeff_1} and \eqref{eq:composition_coeff_2} are nontrivial expressions, it is straightforward to argue (and explicitly prove) that
\begin{equation}
\label{eq:hermitean_condition}
    \zeta(v, v) = 1,
\end{equation}
since $\mu(v)$ is Hermitian with $\pm 1$ eigenvalues, and therefore satisfies $\mu(v)^2 = I \equiv \mu(0)$. Furthermore, the overall phase simplifies dramatically (and becomes real) in the case of $v^T \symp v' = 0$, which we will now show corresponds to $\majo(v)$ commuting with $\majo(v')$.

\subsection{(Anti-) Commutation}
\label{sec:commutation}

Despite the complicated composition law \eqref{eq:majorana_composition} for two Majorana strings, things simplify considerably if we consider just their commutation:
\begin{equation}
\label{eq:majorana_commutation}
\majo(v) \majo(v') = (-1)^{v^T \symp v'} \cdot \majo(v') \majo(v).
\end{equation}
The aforementioned expression $v^T \symp v'$ therefore indicates if two strings commute or anti-commute. And since any operator commutes with itself, it also has to satisfy
\begin{equation}
\label{eq:symp_self_commuting}
v^T \symp v = 0 \quad \forall v \in \F^{2n}.
\end{equation}
It's straightforward to show that this is indeed the case, e.g.\ using \eqref{eq:symp_acts_on_vector} in Section \ref{sec:weight_parity}. With this property of $\symp$ we can hence make the following statement:

\begin{lemma}[Symplectic Product]

The map $\inner{\cdot}{\cdot}: \F^{2n} \times \F^{2n} \rightarrow \F$ acting according to
\begin{equation}
    \inner{v}{v'} \coloneqq v^T \symp v'
\end{equation}
defines a \emph{symplectic product}, i.e.\ it is bilinear, alternating ($\inner{v}{v} = 0$) and non-degenerate\footnote{By non-degenerate we mean that the matrix does not have zero eigenvalues. This should not be confused with the notion of degenerate eigenvalues in quantum mechanics.}. $\symp$ is therefore called a \emph{symplectic form}.

\begin{proof}
    $\inner{\cdot}{\cdot}$ is trivially bilinear, and alternating because of \eqref{eq:symp_self_commuting}. It is also non-degenerate because the spectrum of $\symp$ in a vector space $\F^N$ of general dimension $N$ is
    \begin{align}
    \begin{split}
        \lambda_1 &= N-1 \mod{2}, \quad e_1 = (1, 1, \ldots, 1)^T \\
        \lambda_{i} &= 1, \quad e_i = (0, \ldots, 0, \underbrace{1, 1}_{\mathclap{\text{$(i-1)$th and $i$th position}}}, 0, \ldots, 0)^T, \quad 1 < i \leq N.
    \end{split}
    \end{align}
    But since we chose $N = 2n$ to be even, we have $\lambda_1 = 1$ and the spectrum is therefore not degenerate.
\end{proof}
\end{lemma}

\subsection{Operator Weight \& Parity}
\label{sec:weight_parity}

As touched upon in Section \ref{sec:introduction}, the concept of parity is very important for systems composed of Majorana fermions. It does not only apply to the states living in the Hilbert space, but also to the operators acting on them. For Majorana strings we hence define the notions of \emph{(operator) weight and parity} in the following way:

\begin{definition}[Operator Weight \& Parity]
\hfill
\begin{itemize}
    \item The \emph{(operator) weight} $\weight(\majo(v)) \equiv \weight(v)$ of a Majorana string $\majo(v)$ is defined as the number of (elementary) non-identity Majorana operators $\chi_i$ it is composed of. Equivalently, the weight is equal to the number of non-zero entries in the string's vector representation $v \in \F^{2n}$.

    \item  The \emph{(operator) parity} $\parity(\majo(v)) \equiv \parity(v)$ of a Majorana string $\majo(v)$ can be defined as the string's operator weight mod 2. Equivalently, the parity can be defined in terms of the dot product of the string's vector representation with itself or $j$ \eqref{eq:all_ones_vector} (over $\F$):
    \begin{equation}
    \label{eq:parity_def}
        \parity(v) \equiv v^T v = v^T j.
    \end{equation}
    
    If $\parity(v) = 1$, $\majo(v)$ (or $v$) is said to have \emph{odd} parity. If $\parity(v) = 0$, $\majo(v)$ (or $v$) is said to have \emph{even} parity.
    \end{itemize}
\end{definition}

As alluded to before, the observable distinguishing between operators and states of different parities is the aptly named \emph{parity operator}
\begin{equation}
\label{eq:parity_check}
    (-1)^F \equiv \mu(j),
\end{equation}
which is uniquely defined up to an overall sign. Superselection then dictates that any Majorana string corresponding to a physical observable must commute with $(-1)^F$. However, using the general property
\begin{equation}
\label{eq:symp_acts_on_vector}
    \symp v = \parity(v) \cdot j - v = \begin{cases}
        v, & \parity(v) = 0 \\
        v^c, & \parity(v) = 1
    \end{cases},
\end{equation}
one comes to the conclusion that $v^T \symp j = v^T j = \parity(v)$ for all $v \in \F^{2n}$ per \eqref{eq:parity_def}, and thus the the commutation property \eqref{eq:majorana_commutation} implies
\begin{equation}
    (-1)^F \majo(v) \, (-1)^F = (-1)^{\parity(v)} \cdot \majo(v).
\end{equation}
Hence only strings with even parity commute with $(-1)^F$ and can therefore be considered valid observables. This does however not imply that odd-parity strings are without use, as they can for example describe particle insertions or logical operators acting on stabilizer states. The latter is discussed in more detail in Section \ref{sec:logical_operators}.

Besides their physical relevance, even Majorana strings also have the useful property that their parity is preserved under composition with other even strings. Hence it makes sense to identify the subspace of all even-parity vectors in $\F^{2n}$:
\begin{equation}
    \F^{2n, +} \coloneqq \{ v \in \F^{2n} | \parity(v) = 0\},
\end{equation}
and the associated even-parity Majorana subgroup:
\begin{equation}
\label{eq:majorana_parity_group}
    \Majo^+_{2n} \coloneqq \{ (i)^a \cdot \majo(v) \, | \, a \in \Z, v \in \F^{2n, +} \} \leq \Majo_{2n}.
\end{equation}
One can easily see that the same can't be done for Majorana strings with odd parity due to it not being invariant under composition. Nevertheless, it is useful to define their respective counterparts as the \emph{sets} $\F^{2n,-}$ and $\Majo_{2n}^-$.

\subsection{Connection to Pauli Strings}
\label{sec:pauli_strings}

Let's connect what we have established so far to the analogous representation of (qubit) Pauli strings. In that context a binary vector of length $2n$ can be mapped to a tensor product of $n$ elementary Pauli operators \cite{gross_hudsons_theorem_finite_2006}:
\begin{equation}
\label{eq:pauli_string}
    \pauli(v) = (i)^{v^T \symppauliL v} \, X^{v_1} Z^{v_2} \otimes \cdots \otimes X^{v_{2n-1}} Z^{v_{2n}},
\end{equation}
where $X$ and $Z$ are the standard Pauli matrices and $\symppauliL$ is the equivalent of $\sympL$, here corresponding to the lower-triangular part of the matrix
\begin{equation}
\label{eq:symp_matrix_pauli}
    \symppauli = \begin{pmatrix}
            0 & 1 \\ 1 & 0
        \end{pmatrix}^{\oplus \, n}.
\end{equation}
Analogous to $\symp$, $\eta$ encodes the commutation structure of Pauli strings and hence also defines a symplectic product. We can therefore find an invertible linear transformation $W \in \GL(2n, \F)$ that maps between these two symplectic forms:
\begin{equation}
\label{eq:symp_prod_equiv}
    \omega = W^T \, \eta \, W.
\end{equation}
This has the implication that if two Majorana strings $\majo(v), \majo(v')$ (anti-)commute, then the corresponding Pauli strings $\pi(W v), \pi(W v')$ also do:
\begin{equation}
    \inner{v}{v'}_{\majo} = \inner{W v}{W v'}_{\pauli}
\end{equation}
What $W$ therefore describes is the $\F$-representation of a \emph{Jordan-Wigner} transformation between operator product strings of Majorana fermions, and tensor product strings of Pauli operators \cite{jordan_ueber_paulische_aequivalenzverbot_1928}:
\begin{equation}
\label{eq:majo_pauli_equiv}
    \majo(v) \cong \pauli(W v).
\end{equation}
This is trivially true for $n = 1$ since in that case $\symp = \symppauli$, which coincides with the natural identification $X \cong \chi_1$, $Z \cong \chi_2$. In general for our choice of basis, $W$ is of the form
\begin{equation}
\label{eq:jordan_wigner_matrix}
    W = (\symppauli^c)_U = \begin{pmatrix}
        1 & 0 & 1 & 1 & \cdots & 1 & 1 \\
        0 & 1 & 1 & 1 & \cdots & 1 & 1 \\
        0 & 0 & 1 & 0 & \cdots & 1 & 1 \\
        0 & 0 & 0 & 1 & \cdots & 1 & 1 \\
        \vdots & \vdots & \vdots & \vdots & \ddots & \vdots & \vdots \\
        0 & 0 & 0 & 0 & \cdots & 1 & 0 \\
        0 & 0 & 0 & 0 & \cdots & 0 & 1 \\
    \end{pmatrix},
\end{equation}
which is the upper-triangular part of the matrix complement of $\symppauli$ (including the diagonal). To prove this, one first has to confirm that
\begin{equation}
\label{eq:symp_pauli_LU_decomp}
    \symppauli^c = (\symppauli^c)_L (\symppauli^c)_U,
\end{equation}
which can be done e.g.\ by induction over $n$. Inserting $I = \symppauli^2$ and applying \eqref{eq:symp_acts_on_vector} to the odd-parity rows/columns of $\symppauli$ then yields
\begin{equation}
    \symp = \symppauli \symppauli \symp = \symppauli \symppauli^c = \symppauli \, (\symppauli^c)_L (\symppauli^c)_U = (\symppauli^c)_L \symppauli \,(\symppauli^c)_U,
\end{equation}
where in the last step we also used the fact that $\symppauli$ and $(\symppauli^c)_L$ commute. Additionally, \eqref{eq:symp_pauli_LU_decomp} implies that one can also go the other way, as inserting $\symp$ (which commutes with all matrices whose rows and columns have the same parity) into that expression results in
\begin{equation}
    \symppauli = (\symppauli^c)_L \symp \,(\symppauli^c)_U.
\end{equation}
In fact, one can prove (again using induction) that
\begin{equation}
    W^2 = I,
\end{equation}
which means that $W$ both maps vector representations of Majorana strings to ones of Pauli strings, and vice versa. From the structure of \eqref{eq:jordan_wigner_matrix} it also becomes apparent that low-weight strings in one basis are generally mapped to high-weight strings in the other, which illustrates the non-locality of the transformation. This is to be expected, as the most basic Jordan-Wigner transformations are used to assign local fermionic degrees of freedom to non-local topological domain walls, e.g.\ in the one-dimensional (bosonic) XY model \cite{lieb_two_soluble_models_1961}.

The matrix $W$ does however not completely fix the corresponding Jordan-Wigner transformation, as it is agnostic with regard to specific sign choices for the operators. A canonical mapping with regards to our conventions is therefore provided in Appendix \ref{sec:canonical_jordan_wigner}. However, in the presence of a (bosonic) Hamiltonian $H$ it might be necessary to make a completely different choice for $W$ that complies with the structure of $H$, or even use a different type of transformation altogether \cite{elman_free_fermions_disguise_2021}. The ramifications of this for our framework could be explored in future work. 

\subsection{Hilbert Space Representation}

So far we have not mentioned a concrete representation of Majorana strings on a Hilbert space $\Hilbert$. Fortunately, thanks to the identification with Pauli strings we can simply choose the one defined implicitly in terms of \eqref{eq:majo_pauli_equiv} and \eqref{eq:pauli_string}. This means that the trace of a Majorana string (also referred to as the \emph{character} of the chosen representation) is proportional to
\begin{equation}
\label{eq:majo_trace}
    \tr(\majo(v)) = \tr(\pi(Wv)) =  2^n \delta_{v, 0},
\end{equation}
i.e.\ all strings except for multiples of the identity are traceless, and the dimension of $\Hilbert$ is $2^n$.

\begin{lemma}[Irreducibility]
\label{thm:majo_irreducible}
    The Hilbert space representation of Majorana strings defined in terms of \eqref{eq:majo_pauli_equiv} and \eqref{eq:pauli_string} is irreducible.
    \begin{proof}
        It is a known result of representation theory (see e.g.\ \cite{simon_representations_finite_compact_1996}) that it suffices for the averaged sum over the absolute square of the characters $\tr((i)^a \majo(v))$ to be equal to 1, which is indeed the case due to \eqref{eq:majo_trace}:
        \begin{equation}
            \frac{1}{|\Z|} \frac{1}{|\F^{2n}|} \sum_{a \in \Z} \sum_{v \in F^{2n}} |\tr((i)^a \majo(v))|^2 = 1.
        \end{equation}
    \end{proof}
\end{lemma}
While taking the trace of a Majorana string (or any operator that can be expanded in terms of Majorana strings) is straightforward, the lack of a tensor product structure in \eqref{eq:majo_string} means that the Hilbert space has no obvious factorization with regard to the individual mode operators $\chi_i$, and hence no way to take a partial trace. Instead one has to adopt the notion of \emph{Majorana (sub-) algebras} \cite{benatti_entanglement_algebraic_quantum_2016}. We provide the basics of this formalism in Appendix \ref{sec:majorana_algebras} due to its importance e.g.\ for analyzing the entanglement and entropy of fermionic subsystems.

\section{The Majorana Clifford Group}
\label{sec:cliffords}

Having laid out the basic principles of Majorana strings, it is now time to consider the possible ways to transform (between) them. Quantum mechanics dictates that any such transformation should take the form of a unitary (super-) operator, and the largest class of such operators that map Majorana strings to other Majorana strings is called the \emph{(Majorana) Clifford group}:

\begin{definition}[Majorana Clifford Group]
    \label{def:clifford_group}
    The \emph{Majorana Clifford group} $\Cliff_{2n}$ is the group of all unitary automorphisms acting on the Majorana group $\Majo_{2n}$ \eqref{eq:majorana_group}, i.e.\ all unitary operators $\cliff$ that map (general) Majorana strings to other such strings (up to a phase):
    \begin{equation}
    \label{eq:clifford_group_def}
        \cliff \majo(v) \cliff^\dagger \equiv c(v) \cdot \majo(S(v))
    \end{equation}
    for some functions $c: \F^{2n} \rightarrow U(1)$ and $S: \F^{2n} \rightarrow \F^{2n}$ that depend on $\cliff$.
\end{definition}

Understanding the structure of the Clifford group is therefore equivalent to understanding the allowed options for $S$ (and to lesser extent $c$).

\subsection{Symplectic Structure}

Unitarity generally preserves the (anti-) commutation relations of operators $O_1, O_2$ in the sense that
\begin{equation}
    [U O_1 U^\dagger, U O_2, U^\dagger] = U [O_1, O_2] U^\dagger.
\end{equation}
The same must therefore be true for elements of the Clifford group. Since it was shown in section \ref{sec:commutation} that two Majorana strings $\majo(v), \majo(v')$ (anti-) commuting is equivalent to the symplectic product $\inner{v}{v'}$ being zero (or one), the function $S$ associated to a Clifford element as described in definition \ref{def:clifford_group} has to preserve the structure of said product. We later prove in Theorem \ref{thm:clifford_symplectic} that any $S$ that does so is necessarily linear and hence \emph{symplectic}:

\begin{definition}[Symplectic Group]
\label{def:symplectic}
    The (binary) \emph{symplectic group} $\Sp(2n) \equiv \Sp(2n, \F)$ is the group of all invertible linear
transformations $S \in \GL(2n, \F)$ that preserve the symplectic product:
\begin{equation}
    \label{eq:symp_group_def}
        \inner{S v}{S v'} = \inner{v}{v'} \quad \forall \, v,v' \in \F^{2n}.
    \end{equation}
\end{definition}

This definition does not depend on any specific choice of basis, meaning it applies equally to the Majorana and Pauli case. In the basis of former, \eqref{eq:symp_group_def} is equivalent to
\begin{equation}
\label{eq:symp_group_def_base}
    S^T \omega S = \omega \quad \forall S \in \Sp(2n).
\end{equation}
Obtaining the associated group element in the Pauli basis (where one would require $S^T \eta S = \eta$ instead) is then as simple as applying a Jordan-Wigner transformation \eqref{eq:jordan_wigner_matrix}:
\begin{equation}
    S \rightarrow W S W.
\end{equation}

\begin{proposition}[Group Order of $\Sp(2n)$]
    The total number of elements of the symplectic group $\Sp(2n)$ is \cite{grove_classical_groups_geometric_2002}
    \begin{equation}
    \label{eq:symp_group_order}
        \lvert \Sp(2n) \rvert = 2^{n^2} \prod_{i=1}^n (4^i - 1).
    \end{equation}
\end{proposition}

A method to uniformly and efficiently sample a symplectic matrix $S_i$ (in the Pauli basis) by associating it to a unique integer $1 \leq i \leq \lvert \Sp(2n) \rvert$ is illustrated in \cite{koenig_how_efficiently_select_2014}.

With that we can formalize and further expand upon the previous statements to fully classify the elements of the (Majorana) Clifford group $\Cliff_{2n}$:

\begin{theorem}[Symplectic Representation of Cliffords]
\label{thm:clifford_symplectic}
\hfill
\begin{enumerate}
    \item For $\cliff$ as in \eqref{eq:clifford_group_def} to be an element of the Majorana Clifford group $\Cliff_{2n}$, the associated mapping $S$ must be linear and therefore symplectic. Conversely, for any symplectic matrix $S$ there is at least one $\cliff \in \Cliff_{2n}$ that acts accordingly.

    \item If $\cliff(S)$ is a Clifford with associated symplectic $S$, then (ignoring overall phases) $\majo(a) \cliff(S)$, $a \in \F^{2n}$ is also a Clifford with the same associated symplectic $S$. Conversely, any two Clifford group elements $\cliff(S), \cliff'(S)$ are equivalent up to some Majorana string $\majo(a)$ and a phase:
    \begin{equation}
        \cliff(S) \sim \cliff'(S) = e^{i \phi} \majo(a) \cliff(S).
    \end{equation}

    \item Let $[\cliff]: \Sp(2n) \rightarrow \Cliff_{2n} / \sim$ be the mapping that assigns to each symplectic $S$ the corresponding equivalence class $[\cliff(S)]$ of Cliffords. Then this mapping defines a representation of $\Sp(2n)$:
    \begin{equation}
        [\cliff(S)] \, [\cliff(T)] = [\cliff(S T)].
    \end{equation}

\end{enumerate}
\end{theorem}

A proof for this discrete variation of the \emph{Stone-von Neumann Theorem}\footnote{We call it a Stone-von Neumann theorem in accordance with the equivalent theorem for Pauli Cliffords \cite{gross_hudsons_theorem_finite_2006}. There one can establish an explicit connection between the commutation relations of $Z$ and $X$, and the commutation relations of position and momentum operators.} is found in Appendix \ref{sec:stone_neumann_proof}.

Note that it is not possible to find a closed form expression for the coefficients $c(v)$ in \eqref{eq:clifford_group_def}. This is again due to the nontrivial composition rules \eqref{eq:majorana_composition} of two Majorana strings. Still, by evaluating $\cliff(S) \majo(v) \majo(v') \cliff^\dagger(S)$ in two different ways one can obtain the following relation:
\begin{equation}
\label{eq:cliff_coeff_composition}
    c(v) \cdot c(v') = \frac{\zeta(v, v')}{\zeta(Sv, Sv')} \cdot c(v + v'),
\end{equation}
which implies that $c(v)^2 = c(0) = 1$ due to \eqref{eq:hermitean_condition} and $\majo(0) = I$. Hence one has
\begin{equation}
\label{eq:clifford_phase_pm}
    c(v) = \pm 1 \quad \forall v \in \F^{2n},
\end{equation}
as expected from unitarity. Stronger statements can be made for qudit Pauli strings with $d > 2$ prime \cite{gross_hudsons_theorem_finite_2006}, but since the overall phase of a quantum state has no physical meaning\footnote{The phases do however become important when considering superpositions of Majorana strings or quantum states, in which case a more detailed analysis is necessary. This could be explored in future work.}, we can safely ignore it most of the time and focus solely on the symplectic representation $S$. $\cliff(S)$ is therefore used to denote any Clifford unitary that acts accordingly.

\begin{corollary}[Order of the Clifford Group]
\label{thm:clifford_order}
The size of the quotient set $\Cliff_{2n} / \Unit(1)$, i.e.\ the number of elements in $\Cliff_{2n}$ that (ignoring phases) are pairwise distinct, is
\begin{equation}
    \lvert \Cliff_{2n} / \Unit(1) \rvert = 2^{n^2 + 2n} \prod_{i=1}^n (4^i - 1).
\end{equation}
\begin{proof}
    Theorem \ref{thm:clifford_symplectic} directly implies that the quotient set size factorizes according to 
    \begin{equation}
        |\Cliff_{2n} / \Unit(1)| = |\Sp(2n)| \cdot |\F^{2n}|.
    \end{equation}
    Plugging in \eqref{eq:symp_group_order} and $|\F^{2n}| = 2^{2n}$ then results in what was to be shown. 
\end{proof}
\end{corollary}

\subsection{Parity-Preserving Cliffords}

So far the Pauli and Majorana Clifford groups are mathematically equivalent. But unlike Pauli strings, Majorana strings have the additional extrinsic property of operator parity. It is easy to see that Clifford operators do not preserve thar parity in general, e.g.\
\begin{equation}
    \begin{pmatrix}
        1 & 0 \\ 1 & 1
    \end{pmatrix}
    \begin{pmatrix}
        1 \\ 0
    \end{pmatrix}
    = \begin{pmatrix}
        1 \\ 1
    \end{pmatrix}
\end{equation}
depicts a symplectic matrix (Clifford operator) mapping between vectors (Majorana strings) of different parity. This can be interpreted as observables not being mapped to other observables under such transformations, which is not physically reasonable.

For an element $\cliff(S)$ to respect the parity of any string, the associated symplectic $S$ must satisfy
\begin{equation}
    p(S v) = (S v)^T (S v) = v^T (S^T S) v \stackrel{!}{=} v^T v = p(v)
\end{equation}
for all $v \in \F^{2n}$, and therefore be \emph{orthogonal}:

\begin{definition}[Orthogonal Group]
\label{def:orthogonal_group}
    The (binary) \emph{orthogonal group} $\Ortho(2n) \equiv \Ortho(2n, \F)$ is the group of all invertible linear transformations $A \in \GL(2n, \F)$ that preserve the canonical inner product:
\begin{equation}
    \label{eq:ortho_group_def}
        A^T A = A A^T = I.
    \end{equation}
\end{definition}

In fact, any orthogonal matrix is also symplectic, meaning that $\Ortho(2n)$ is a subgroup of the symplectic group $\Sp(2n)$\footnote{In the Majorana basis this is easy to see, as any $A \in \Ortho(2n)$ can only have odd-parity row and column vectors, leading to
\begin{equation}
     A^T \symp A = A^T A^c = A^T A \symp = \symp.
\end{equation}}:
\begin{equation}
    \Ortho(2n) \leq \Sp(2n).
\end{equation}
This also implies that the order of $\Ortho(2n)$ must divide the order of $\Sp(2n)$ \eqref{eq:symp_group_order}, which is true since:

\begin{proposition}[Group Order of $\Ortho(2n)$]
    The total number of elements of the orthogonal group $\Ortho(2n)$ is \cite{grove_classical_groups_geometric_2002}
    \begin{align}
    \begin{split}
    \label{eq:ortho_group_order}
        \lvert \Ortho(2n) \rvert &= 2^{n^2} \prod_{i=1}^{n-1} (4^i - 1) \\
        &= \lvert \Sp(2n) \rvert / (4^n - 1).
    \end{split}
    \end{align}
\end{proposition}

We can therefore identify a subgroup of the Clifford group $\Cliff_{2n}$ that acts according to $\Ortho(2n)$ and thus preserves parity:

\begin{corollary}(Parity-Preserving Clifford Subgroup)
    The Majorana Clifford group $\Cliff_{2n}$ admits a parity-preserving subgroup $\Cliff^{\parity}_{2n}$. Its elements are of the form $\majo(a) \cliff(S)$ with $S \in \Ortho(2n)$ and $a \in \F^{2n, +}$. The size (mod phases) of this so-called \emph{p-Clifford group} is therefore
    \begin{align}
    \begin{split}
        |\Cliff^{\parity}_{2n} / \Unit(1)| &= |\Ortho(2n)| \cdot |\F^{2n, +} | \\
        &= 2^{n^2 + 2n - 1} \prod_{i=1}^{n-1} (4^i - 1).
    \end{split}
    \end{align}
\end{corollary}
Note that we have $\Majo_{2n}^+ \leq \Cliff^{\parity}_{2n}$, because while odd-parity strings don't affect the operator parity of other strings when acting from both sides, they can indeed change the parity of states in the Hilbert space. This is shown explicitly in Section \ref{sec:stabilizer_space}. Overall, all elements of $\Cliff^{\parity}_{2n}$ hence commute with $(-1)^F \equiv \majo(j)$ \eqref{eq:parity_check} because
\begin{equation}
\label{eq:j_vector_invariant}
    \cliff(S) \majo(j) \cliff^\dagger(S) = c(j) \cdot \majo(S j) = \majo(j),
\end{equation}
where $S j = j$ due to all $S \in \Ortho(2n)$ having odd-parity rows, and $c(j) = 1$ using arguments for fixed points of $S$ made in Appendix \ref{sec:parity_frame_proof}.

\subsection{Generating Set \& Unitary Representation}
\label{sec:generating_set}

In the Pauli basis, the \emph{Gottesman-Knill theorem} \cite{gottesman_heisenberg_representation_quantum_1998} states that the Clifford group $\Cliff_{2n}$ can be efficiently generated from repeated applications of tensor-products containing CNOT, Hadamard, and the phase gate (or an equivalent minimal set) as the only non-trivial operators. In the symplectic representation, this is equivalent to the fact that any element $S \in \Sp(2n)$ can be decomposed into a product of symplectic \emph{transvections}
\begin{align}
\begin{split}
\label{eq:transvection}
    & \tau_a: v \mapsto v + \inner{a}{v}_{\pauli} a \\
    \Longleftrightarrow \quad & \tau_a = I + a \, a^T \symppauli, \quad \quad \quad a \in \F^{2n}.
\end{split}
\end{align}

Additionally, all such transvections can be written as products and (in the Pauli basis) direct sums of just 3 independent ones:
\begin{itemize}
    \item For $n = 1$, there are two independent transvections (corresponding to e.g.\ $s = (0, 1)^T$ and $h = (1, 1)^T$), from which all other symplectics can be generated by repeated multiplication of $\tau_s$ (which represents the phase gate) and $\tau_h$ (which represents the Hadamard gate).

    \item For $n = 2$, it can be shown that there is only one independent transvection $\tau_c$ (for e.g.\ $c = (0, 1, 1, 0)^T$) since any other one can be written as $\tau_c$ times some direct sum of $n = 1$ symplectics. In fact, CNOT corresponds to the symplectic $(\tau_s \oplus \tau_h \tau_s \tau_h) \, \tau_c$ in the Pauli basis.

    \item For $n > 2$ no other independent transvections exist.
\end{itemize}

It is therefore reasonable to assume that similar statements also hold for $\Cliff_{2n}$ in the Majorana basis, and indeed there too the symplectic group is generated by transvections with regard to $\symp$:
\begin{align}
\begin{split}
    W \tau_a W &= I + (W a) (W a)^T \symp \\
    &\equiv h_{(W a)}.
\end{split}
\end{align}

The unitary representations of transvections in the Majorana basis are usually referred to as \emph{braiding operators}, and can be defined as follows:

\begin{definition}[Braiding Operators] 
    For any Hermitian Majorana string $\majo(a)$ with $a \in \F^{2n}$, the associated braiding operator $\braid(a) \equiv \braid(\majo(a))$ is defined as
    \begin{align}
    \label{eq:braid_operators}
    \begin{split}
        \braid(a) &\coloneqq \exp\left(i \frac{\pi}{4} \majo(a) \right) \\
        &= \frac{1}{\sqrt{2}} \big(I + i \majo(a)\big).
    \end{split}
    \end{align}
\end{definition}
That these operators indeed act as symplectic trans\-vections is straightforward to prove by letting them act on a Majorana string:
\begin{align}
\begin{split}
\label{eq:braiding_action}
    & \braid(a) \majo(v) \braid^\dagger(a) \\
    = {}& \frac{1}{2} \left(1 + (-1)^{\inner{a}{v}}\right) \majo(v) \\
    &+ \frac{i}{2} \left(1 - (-1)^{\inner{a}{v}}\right) \majo(a) \majo(v) \\
    \equiv {}& (i \, \zeta(a, v))^{\inner{a}{v}} \cdot \majo(v + \inner{a}{v} a),
\end{split}
\end{align}
where the last line can be derived by comparing the cases of $\inner{a}{v} = 0$ and $\inner{a}{v} = 1$. We can therefore identify
\begin{align}
\begin{split}
    \cliff(h_a) &\sim \braid(a), \\
    \quad c(v) &\sim (i \, \zeta(a, v))^{\inner{a}{v}} \\
    &= (-1)^{\inner{a}{v} (a^T \sympL v + f(a, v) + 1)},
\end{split}
\end{align}
with the latter equality following from \eqref{eq:composition_coeff_1}, thereby making the validity of \eqref{eq:clifford_phase_pm} manifest. Note that $\braid^\dagger(a)$ acts in the exact same way, except for sign change according to $c(v) \rightarrow (-1)^{\inner{a}{v}} c(v)$.

However, unlike in the Pauli case, direct sums of transvections in the Majorana basis (or equivalently fermionic tensor products of braiding operators) are not necessarily valid transvections anymore. In fact, it is only possible for transvections $h_a$ with $a \in \F^{2n, +}$\footnote{This can be shown e.g.\ by writing a generic direct sum of transvections as a block matrix and checking under which conditions it still satisfies \eqref{eq:symp_group_def_base}.}, in which case \eqref{eq:symp_acts_on_vector} dictates that they must be of the form
\begin{align}
\begin{split}
\label{eq:householder}
    & h_a: v \mapsto v + (a^T v) \, a \\
    \Longleftrightarrow \quad & h_a = I + a \, a^T, \quad \quad \quad a \in \F^{2n, +}.
\end{split}
\end{align}
Such transformations are called \emph{Householder reflections}, and it should come as no surprise that they correspond to the parity-preserving subset of symplectic transvections. In the style of Gottesman-Knill, one can then prove that the corresponding set of braiding operators (efficiently) generates $\Cliff^{\parity}_{2n} / \Majo^+_{2n}$:

\begin{theorem}[Gottesman-Knill for p-Cliffords]
\label{thm:gottesman_knill}
    Any element of the p-Clifford group $\Cliff^{\parity}_{2n} \leq \Cliff_{2n}$ (mod $\Majo^+_{2n}$) can be efficiently generated from products of braiding operators $\braid(a)$ with $a\in \F^{2n,+}$ as in \eqref{eq:braid_operators}. More specifically, it can be efficiently generated from the subset of braiding operators satisfying $\weight(a) = 2$, $\weight(a) = 4$ or $a=j$ (compare to \cite{mclauchlan_fermion_parity_based_2022, mudassar_encoding_majorana_codes_2024}). 
\begin{proof}
    We first note the well-known fact that the (binary) orthogonal group $\Ortho(2n)$ is generated by the set of Householder reflections as defined in \eqref{eq:householder} \cite{grove_classical_groups_geometric_2002}. This can be done in $\mathcal{O}(n^3)$ steps, as demonstrated by the sampling algorithm we provide in appendix \ref{sec:sampling}. Using \eqref{eq:braiding_action}, it is then straightforward to argue that the Householder reflections efficiently generate the orthogonal action of $\Cliff^{\parity}_{2n}$. 
    
    What is left to show is that all reflections (except $h_j \equiv \omega$) can be generated from just the ones corresponding to vectors with weight 2 and weight 4. This can be done in at most $\mathcal{O}(n^3)$ steps by first noting that the Householder reflections exhibit a braiding algebra:
    \begin{equation}
    \label{eq:reflection_braiding}
        h_b h_a h_b = h_{h_b a}.
    \end{equation}
    Given some reflection associated to a vector $a$ with weight $k > 4$ even, the problem therefore reduces to finding weight-4 reflections that map it to a vector with weight 2 or 4. This is always possible because any weight-4 Householder is of the form $\symp_4 \oplus I_{2n-4}$ up to permutations (which correspond to weight-2 reflections). This essentially allows us to use \eqref{eq:symp_acts_on_vector} with regard to any 4 entries of $a$ with local weight 3 to map them to their weight-1 complement, thereby reducing the overall weight by 2. For example, locally:
    \begin{equation}
        \symp h_{(0, 1, 1, 1)^T} \symp = h_{\symp (0, 1, 1, 1)^T} = h_{(1, 0, 0, 0)^T} 
    \end{equation}
    
    This reduction has to be performed $\mathcal{O}(k)$ times, each step involving two $\mathcal{O}(n^2)$ matrix applications due to \eqref{eq:reflection_braiding}.
\end{proof}
\end{theorem}

Even though the nonlocality of most braiding operations (and by extend Cliffords) makes their behavior unintuitive compared to their qubit counterparts, it is worth emphasizing that operations corresponding to Householder reflections with weight 2 act as simple permutation operators. This is further elaborated on in Appendix \ref{sec:fermionic_swap}.

\section{Fermionic Stabilizers}
\label{sec:stabilizers}

We now discuss how to define and construct general fermionic stabilizers in terms of Majorana operator strings, such that obey the parity superselection rules.

\begin{definition}[Fermionic Stabilizer]
    A $[[n, k]]_f$ (Majorana) fermionic stabilizer $\Stab$ is an abelian subgroup of the even-parity Majorana group $\Majo^+_{2n}$ that is generated by $r = n-k \leq n$ independent strings and does not contain the operator $-I$. 
\end{definition}

A simple example is provided in Appendix \ref{sec:shortest_fermion_code}, and further such stabilizers can be found in e.g.\ \cite{bravyi_majorana_fermion_codes_2010, mudassar_encoding_majorana_codes_2024, vijay_quantum_error_correction_2017, litinski_quantum_computing_majorana_2018, vijay_majorana_fermion_surface_2015}. 

\subsection{Symplectic Representation}

The reason why a stabilizer can have at most $n$ independent generators becomes apparent when one considers its binary equivalent:

\begin{definition}[Isotropic Subspaces] A subspace $M$ of the vector space $\F^{2n}$ is said to be \emph{isotropic} iff
\begin{equation}
    \inner{m}{m'} = 0 \quad \forall m, m' \in M.
\end{equation}
The largest possible isotropic subspaces have dimension $n$, and are called \emph{maximally isotropic} \cite{gross_hudsons_theorem_finite_2006}.
\end{definition}
Due to fermion superselection we only care about isotropic subspaces of the even-parity subspace $\F^{2n, +}$. In this case any maximally isotropic subspace can be written as\footnote{Note that the orthogonal matrix $S$ is not at all unique.}
\begin{equation}
\label{eq:max_isotropic_general}
    M_\text{max} \equiv S M_\text{std}, \quad S \in \Ortho(2n),
\end{equation}
where $M_\text{std}$ is the canonical maximally isotropic subspace, here provided in terms of a $n \times 2n$ matrix with the basis elements as row vectors:
\begin{equation}
\label{eq:std_max_isotropic_basis}
    M_{\text{std}} \sim \begin{pmatrix}
        1 & 1 & 0 & 0 & 0 & \cdots & 0 & 0 & 0 \\
        0 & 0 & 1 & 1 & 0 & \cdots & 0 & 0 & 0 \\
        \vdots & \vdots & \vdots & \vdots & \vdots & \ddots & \vdots & \vdots & \vdots \\
        0 & 0 & 0 & 0 & 0 & \cdots & 0 & 1 & 1
    \end{pmatrix}.
\end{equation}
It is easy to check that one can't add more rows that are orthogonal with regard to the already present ones.

Each element of an (even-parity) isotropic subspace $M \subset \F^{2n, +}$ of dimension $r = n - k$ can be associated to a Majorana string of (at least one) $[[n, k]]_f$ fermionic stabilizer $\Stab$. Fixing the basis of $M$ to be $\{b_1, \ldots, b_k\}$, one possible choice for this mapping is given by
\begin{equation}
    \Stab \sim \majoBar(M) \coloneqq \{ \majoBar(m) \, | \,  m \in M \},
\end{equation}
where we defined 
\begin{equation}
    \majoBar(m) \coloneqq \prod_{i = 1}^k \majo(b_i)^{m_i}, \quad m = \sum_{i = 1}^k m_i b_i \in M.
\end{equation}
Using the modified (Hermitian) Majorana string $\majoBar(m)$ instead of $\majo(m)$ is beneficial because the resulting effective phases factors of the stabilizer strings are compatible with composition:
\begin{equation}
    \majoBar(m) \majoBar(m') = \majoBar(m + m') \quad \forall m,m' \in M,
\end{equation}
which is only possible due to the strings $\majo(b_i)$ commuting pairwise. Any other possible stabilizer associated to $M$ differs from $\majoBar(M)$ only by the signs in front of each element $\majoBar(m)$ \cite{gross_hudsons_theorem_finite_2006}. These signs naturally must also be compatible with composition (i.e.\ form a \emph{character}). More precisely:

\begin{theorem}
    Any $[[n, k]]_f$ (Majorana) fermionic stabilizer $\Stab$ is of the form
    \begin{equation}
        \Stab(M, v) = \{ (-1)^{v^T m} \majoBar(m) \, | \, m \in M \},
    \end{equation}
    where $M \subset \F^{2n, +}$ is an even-parity isotropic subspace of dimension $r = n - k$, and $v$ is an element of the quotient set $\F^{2n} / M$ \footnote{$v$ is an element of the quotient space because $v$ and $v + m$ with $m \in M$ correspond to the same character.}.
\end{theorem}
A priori, we allow $v$ to have odd parity, as it only affects the leading phases. We will return to this in section \ref{sec:stabilizer_space}.

\subsection{Stabilizer Spaces \& States}
\label{sec:stabilizer_space}

Using the elements of a stabilizer $\Stab(M, v)$, we can construct a projection operator onto the subspace of \emph{logical states}, i.e.\ those states that have unit eigenvalue:

\begin{lemma}
    Given a $[[n, k]]_f$ (Majorana) fermionic stabilizer $\Stab(M, v)$, then the space of states $\ket{\psi} \in \Hilbert$ satisfying 
    \begin{equation}
    \label{eq:stabilizer_action}
        (-1)^{v^T m} \majoBar(m) \ket{\psi} = \ket{\psi} \quad \forall m \in M
    \end{equation}
    is $2^k$-dimensional, and projected onto by the operator
    \begin{equation}
        \Pi(M, v) = \frac{1}{|M|} \sum_{m \in M} (-1)^{v^T m} \majoBar(m).
    \end{equation}
    \begin{proof}
        It is straightforward to check that $\Pi(M, v)^2 = \Pi(M, v)$ due to $M$ forming a group under addition. Hence the dimension of the space it projects onto is
        \begin{equation}
            \tr(\Pi(M, v)) = \frac{2^n}{|M|} = 2^{n-r} = 2^k.
        \end{equation}
        For the same reasons, the projection operator satisfies
        \begin{equation}
            (-1)^{v^T m} \majoBar(m) \Pi(M, v) = \Pi(M, v) \quad \forall m \in M,
        \end{equation}
        and therefore only projects onto the logical states.
    \end{proof}
\end{lemma}

In particular, for the case $r=n$ (i.e. $M$ being maximally isotropic), we can identify a single (logical) stabilizer state $\ket{M, v}$ satisfying \eqref{eq:stabilizer_action}, such that
\begin{equation}
    \Pi(M, v) \equiv \ket{M, v}\bra{M, v} , \quad \dim(M) = n.
\end{equation}
The set $\{\ket{M, v}  \, | \, v \in \F^{2n} / M \}$ forms a basis of $\Hilbert$.

Note that even though the projection operators are composed of even-parity stabilizer elements, they don't necessarily project onto Hilbert space states that have even parity. Indeed, the eigenvalue of $(-1)^F \equiv \majoBar(j)$\footnote{We can write $(-1)^F$ as $\majoBar(j)$ because $j$ is guaranteed to be an element of $M$ if it is maximally isotropic.} associated to a stabilizer state $\ket{M, v}$ is
\begin{equation}
    \langle (-1)^F \rangle = \tr(\majoBar(j) \, \ket{M, v}\bra{M, v}) = (-1)^{\parity(v)}.
\end{equation}
Applying a Majorana string $\majo(a)$ to the state so that
\begin{equation}
\label{eq:parity_change}
    \majo(a) \ket{M, v} \bra{M, v} \majo(a) = \ket{M, v + a} \bra{M, v + a}
\end{equation}
therefore changes the parity of the state if $\parity(a) = 1$, even though the parity of the stabilizer elements is unaffected. This further illustrates why only the even-parity Majorana group $\Majo^+_{2n}$ is a subgroup of $\Cliff^{\parity}_{2n}$, and not $\Majo_{2n}$.

\subsection{Logical Operators \& Error Correction}
\label{sec:logical_operators}

When the fermionic stabilizer is not maximal, i.e.\ the space of logical states $\Hilbert_\text{logic}$ has dimension $2^k > 1$, one can identify Majorana string operators that map between those states while respecting the overall structure. This leads to the following definition:

\begin{definition}[Logical Operators]
    The \emph{logical operators} $\Logic$ associated to a fermionic stabilizer $\Stab$ are the unitary elements of $\Majo_{2n} \setminus \Stab$ that commute with all stabilizer elements (but not necessarily among themselves).
\end{definition}

One can convince oneself that $\Logic$ acting on the logical Hilbert space $\Hilbert_\text{logic}$ is generated by $2k$ independent Majorana strings, analogously to how there are $2n$ independent Majorana strings for the physical Hilbert space $\Hilbert$. Furthermore, those logical operators can have odd parity iff $\pm \majoBar(j)$ is not an element of $\Stab$. This is explicitly allowed because logical operators are not necessarily observables and hence don't have to adhere to parity superselection rules. They are therefore allowed to flip the parity of a logical state similarly to \eqref{eq:parity_change}.

The same also holds true for in the context of error correction, in which case parity-flip errors can occur due to physical system interacting with the environment through the localized exchange of an odd number of Majorana modes (typically one).  In the literature this is referred to as \emph{quasiparticle poisoning} \cite{vijay_quantum_error_correction_2017, mudassar_encoding_majorana_codes_2024}, which is thought to be the primary source of errors in systems whose fundamentally Majorana degrees of freedom are sufficiently spatially separated. Fermionic codes are therefore usually constructed such that that they can detect and correct weight-1 errors, meaning they satisfy the Knill-Laflamme condition \cite{knill_theory_quantum_error_2000}:

\begin{theorem}[Knill-Laflamme]
    The set $\mathcal{E}$ of correctable errors of a stabilizer $\Stab$ satisfies
    \begin{equation}
        \braket{\psi_1 | E_1^\dagger E_2 | \psi_2} = \braket{\psi_1 | \psi_2} \cdot O_{E_1 E_2},
    \end{equation}
    for all error operators $E_1, E_2 \in \mathcal{E}$ and logical stabilizer states $\ket{\psi_1}, \ket{\psi_2} \in \Hilbert_\text{logic}$. $O_{E_1 E_2}$ is a constant depending only on $E_1$ and $E_2$.
\end{theorem}
Roughly speaking, this statement implies that for an error $E$ to be correctable there must be a unique combination of basis elements of $\Stab$ that anticommute with $E$. The resulting \emph{error syndrome} can then theoretically be used to find $E$ and undo it.

However these errors can accumulate over time, and if their total weight becomes larger than that of the smallest logical operator, they are not necessarily detectable anymore. One hence defines:

\begin{definition}[Code Distance \& Stabilizer Code]
    In the context of error correction, the \emph{code distance} $\delta$ of a $[[n, k]]_f$ stabilizer corresponds to the weight of its smallest logical operator. This is then referred to as a $[[n, k, \delta]]_f$ fermionic stabilizer code.
\end{definition}

 It can be shown that any error with weight smaller than $\delta$ can be detected, and any error with weight smaller than $\lfloor (\delta + 1) / 2 \rfloor$ can also be corrected \cite{gottesman_stabilizer_codes_quantum_1997}.

\subsection{Stabilizers from p-Cliffords}
\label{sec:stabs_from_cliffords}

As already implicitly stated in \eqref{eq:max_isotropic_general}, any $[[n, 0]]_f$ stabilizer basis can be generated by applying a single p-Clifford $\cliff \in \Cliff^{\parity}_{2n}$ to each of the standard basis strings $\majo(b_i)$ of $M_\text{std}$ in \eqref{eq:std_max_isotropic_basis}. However, things become non-trivial when considering stabilizers with $k > 0$, as those might or might not contain the element $\pm \majoBar(j)$. And since $j$ gets mapped to itself under application of any orthogonal matrix \eqref{eq:j_vector_invariant}, applying a p-Clifford to any stabilizer (not) containing $\pm \majoBar(j)$ can only result in some other stabilizer (not) containing $\pm \majoBar(j)$. This is problematic e.g.\ if one wants to be able to prepare an arbitrary stabilizer using an encoding quantum circuit, or if one wants to uniformly sample over all possible stabilizers numerically.

There are multiple ways to approach this issue, with the one we describe here being inspired by \cite{mudassar_encoding_majorana_codes_2024}, meaning it incorporates additional unoccupied ancilla modes into the stabilizer $\Stab$ to remove any potential presence of $\pm \majoBar(j)$. One can then use (even-parity) braiding operators \eqref{eq:braid_operators} to diagonalize the given basis of $\Stab$ such that it is represented by the first $r$ rows of \eqref{eq:std_max_isotropic_basis}, each with two additional entries representing the ancilla appended at the end. During the process of diagonalization these ancilla modes might be occupied, but one can always choose the braiding operators such that they won't be in the end. The reverse application of these operations (with additional phase changes, which we ignore) hence provides an encoding p-Clifford for $\Stab$ on the extended Hilbert space.

The following algorithm performs this computation in the symplectic representation, i.e. given an ordered basis $(b_1, \ldots, b_r)$ of an isotropic subspace $M \subset \F^{2n,+}$ that does not contain $j$\footnote{We assume that ancilla modes were introduced in advance.}, the routine returns an orthogonal matrix $S \in \Ortho(2n)$ that maps $M^r_\text{std}$ (the isotropic subspace spanned by the first $r$ rows of \eqref{eq:std_max_isotropic_basis}) to $M$, and in particular the standard basis of $M^r_\text{std}$ to $\{b_1, \ldots, b_r\}$.
\begin{figure}[hbt]
\begin{algorithm}[H]
\caption{STABCLIFFORD($M$)}\label{alg:clifford_from_stabilizer}
\begin{algorithmic}
    \Require $M$ as basis $(b_1, \ldots, b_r)$, $j \notin M$
    \Ensure $S \in \Ortho(2n)$ s.th.\ $ S M^r_\text{std} = M$

    \State $S \gets I \in \Ortho(2n)$

    \For{$i = 1,\ldots,r$}
        \State $m \gets S \, b_i$
        \State $m' \gets m_{(2i-1), \ldots, 2n} \in \F^{2(n-i+1),+}$
        \State $e' \gets (1, 1, 0, \ldots, 0) \in \F^{2(n-i+1),+}$
        \If{$m' = e'$}
            \State $a', b' \gets (0, 1, 1, 0, \ldots, 0)^T \in \F^{2(n-i+1),+}$
        \Else
            \State $a', b' \gets \text{HOUSEHOLDER}(e', m')$
        \EndIf

        \State $a \gets m_{1,\ldots,(2i-2)} \oplus a' \in \F^{2n, +}$
        \State $b \gets (0, \ldots, 0)^T \oplus b' \in \F^{2n, +}$
        
        \State $S \gets h_b h_a \, S$
    \EndFor

    \State \Return $S^T$
\end{algorithmic}
\end{algorithm}
\end{figure}
We prove that this algorithm works using induction with regard to the iteration steps:
\begin{enumerate}
    \item Starting with the first basis element $S \, b_1 = b_1$, because of our initial requirements we can always find a set of Householder reflections $h_b h_a$ that map it to the standard basis vector $(1, 1, \ldots, 0)^T$ (see Appendix \ref{sec:sampling}). We therefore add them to $S$ such that they are applied first.

    \item Assuming that $S$ is composed of reflections that map the first $k$ basis elements to their respective standard basis elements, we now have to find a pair of reflections that map $m = S \, b_{k+1}$ to the next standard basis element.
    
    \item Since $m$ must commute with all previous elements, we know that the first $2k-2$ entries of $m$ are composed of consecutive pairs of either ones or zeroes. We also know that the remaining entries can't be all zeros or all ones, because then $m$ would either not be linearly independent, or one would have $j \in \vspan (S \, b_1, \ldots, S \, b_r)$ and therefore also $j \in \vspan (b_1, \ldots, b_r)$, both of which go against the initial assumptions.
    
    \item Let $m'$ be the (even-parity) vector corresponding to those $2(n-i+1)$ trailing entries. We want to find non-identity Householder reflections $h_{a'}, h_{b'}$ that map $(1, 1, 0, \ldots, 0)$ to $m'$, which as mentioned before is always possible.

    \item Using the corresponding even-parity vectors $a' \neq 0$ and $b'$, we can construct even-parity vectors $a$ and $b$ by adding the first $2k-2$ entries of $m$ to the front of $a'$, and adding that many zeros to the front of $b'$. It is straightforward to check that the corresponding reflections $h_a, h_b$ map $m$ to the $(k+1)$th standard basis element, while keeping the previous $k$ elements invariant. One can therefore update $S$  and proceed to the next iteration.

    \item Finally, the inverse/transpose of $S$ is returned.
\end{enumerate}

Since Householder reflections representing the braiding operators can be found and applied to vectors in $\mathcal{O}(n)$ time (see Appendix \ref{sec:sampling}), the total complexity of this algorithm is $\mathcal{O}(n^2r)$. In \cite{mudassar_encoding_majorana_codes_2024}, this procedure is encoded as a quantum circuit which uses the generating set of braiding operators (see Theorem \ref{thm:gottesman_knill}) to perform Gaussian elimination. Both approaches are not unique at all, as multiple distinct Cliffords can embed the same stabilizer.

\section{p-Cliffords as a Unitary Design}
\label{sec:design}

Another property of interest with regard to the parity-preserving Clifford operators is their behavior as a random unitary ensemble. This is a natural thing to investigate, as it is a well-known fact that the complete Clifford group $\Cliff_{2n}$ (in the fundamental representation) forms a \emph{3-design} of the (Haar-random) unitary group $\Unit(2^n)$ \cite{zhu_multiqubit_clifford_groups_2017}:
\begin{definition}[Unitary $t$-Design]
    An ensemble $\ensemble$ of unitary operators acting on a $N$-dimensional Hilbert space $\Hilbert$ is said to be a \emph{$t$-design} iff any ensemble average containing $t$ copies of $U (\cdot) U^\dagger$ matches the corresponding Haar-random average:
    \begin{equation}
        \Exp_{U \in \ensemble} U^{\otimes t} O (U^\dagger)^{\otimes t} = \int d U U^{\otimes t} O (U^\dagger)^{\otimes t}, 
    \end{equation}
    for any linear operator $O$ acting on $\Hilbert^{\otimes t}$.
\end{definition}
From this definition it follows that any $t$-design (with $t > 1$) is necessarily also a $(t-1)$-design, which can be seen by considering $O_t = O_{t-1} \otimes I$.

However, this characterization is hard to work with due to the presence of $O$, so instead we will be making use of the fact that an unitary ensemble $\ensemble$ is a $t$-design iff the associated \emph{$t$-th frame potential}
\begin{equation}
    \Frame_t(\ensemble) \coloneqq \Exp_{U, V \in \ensemble} \, \lvert \tr(U V^\dagger )\rvert^{2t} 
\end{equation}
matches the $t$-th frame potential $\Frame_t(\Unit(N)) \equiv \upsilon(t, N)$ of the unitary group in $N$ dimensions \cite{gross_evenly_distributed_unitaries_2007}. The latter can be shown to have the following relevant properties \cite{zhu_multiqubit_clifford_groups_2017, scott_optimizing_quantum_process_2008}:
\begin{equation}
\label{eq:frame_haar}
    \upsilon(t, N) = \begin{cases}
        \frac{(2t)!}{t! (t+1)!}, & N = 2 \\
        t!, & N \geq t
    \end{cases}.
\end{equation}
In fact, no unitary ensemble can have a frame potential smaller than $\upsilon(t, N)$, which thus provided a lower bound. 

Since we are primarily interested in unitary ensembles that form a group $\mathcal{G}$ under multiplication, we can use that property to simplify $\Frame_t(\mathcal{G})$:
\begin{equation}
    \Frame_t(\mathcal{G}) = \Exp_{U \in \mathcal{G}} \, \lvert \tr U \rvert^{2t}.
\end{equation}
This follows from the fact that $UV^\dagger$ is guaranteed to also be an element of the unitary ensemble in that case. 

Furthermore, Lemma 2 of \cite{zhu_multiqubit_clifford_groups_2017} states that for any subgroup $\mathcal{G} \leq \Cliff_{2n}$ of the Clifford group satisfying $\mathcal{G} \geq \Majo_{2n}$, the frame potential can be expressed in terms of the subgroup's symplectic representation $G \leq \Sp(2n)$:
\begin{equation}
\label{eq:frame_fixed_points}
    \Frame_t(\mathcal{G}) = \frac{1}{|G|} \sum_{S \in G} f(S)^{t-1},
\end{equation}
where $f(S)$ denotes the number of fixed points (i.e.\ unit eigenvectors) of $S$ in $\F^{2n}$. In the language of Majorana strings, this is in part because the absolute trace $|\tr U|^2| \equiv \tr (\majo(a) \gamma(S))|^2$ of a general Clifford is either zero or equal to the number of strings it commutes with.

To deduce from \eqref{eq:frame_fixed_points} the frame potential of a given subgroup $\mathcal{G}$, one can employ computational methods or -- in the case of small $t$ -- the \emph{orbit-stabilizer relation} \cite{zhu_multiqubit_clifford_groups_2017}. It implies that $\Frame_t(\mathcal{G})$ is equivalent to the number of orbits of $G$ acting on $(\F^{2n})^{\times (t - 1)}$, an orbit being a subset of elements that are permuted among each other under the group action. For $t=1,2,3$ this imposes the following requirements for $\mathcal{G}$ to be a $t$-design:
\begin{itemize}
    \item Any subgroup $\mathcal{G}$ is trivially a 1-design as $\Frame_t(\mathcal{G}) = 1 = \upsilon(1, N)$ for all $N$.
    \item $\mathcal{G}$ is a 2-design iff $G$ has 2 orbits on $\F^{2n}$, i.e.\ it is transitive on $\F^{2n} \setminus \{0\}$ since $0 \in \F^{2n}$ is always a fixed point.
    \item $\mathcal{G}$ is a 3-design iff $G$ has 6 orbits on $(\F^{2n})^{\times 2}$, or 5 orbits if $n = 1$ ($N = 2$) due to \eqref{eq:frame_haar},
\end{itemize}
With this once can deduce that $\Cliff_{2n}$ is indeed a 3-design on $\Hilbert$, but $\Cliff^{\parity}_{2n} \leq \Cliff_{2n}$ is \emph{not}, as $\Ortho(2n) \leq \Sp(2n)$ has 4 orbits on $\F^{2n}$:
\begin{equation*}
    \{0\}, \quad \{j\}, \quad \F^{2n, +} \setminus \{0, j\}, \quad \F^{2n, -}.
\end{equation*}
This is a direct consequence of $\Cliff^{\parity}_{2n}$ preserving the parity of Majorana strings, and should therefore not come as a surprise.

\subsection{Fixing the Parity}

However, based on the parity superselection rules we can argue that this is not the correct way to think about $\Cliff^{\parity}_{2n}$ as a $t$-design. The frame potential we are interested in should only care about how p-Cliffords (or unitaries of some other ensemble commuting with $(-1)^F$) act on a fixed-parity sector of $\Hilbert$. Without loss of generality, we choose the even-parity sector for our considerations, and hence define:

\begin{definition}[Parity-Restricted Frame Potential]
    Let $\ensemble$ be an unitary ensemble acting on a Hilbert space $\Hilbert$ such that it preserves fermion parity, i.e.\ $U^\dagger (-1)^F U = (-1)^F$ for all $U \in \ensemble$. Then the corresponding \emph{parity-restricted frame potential} is defined as 
    \begin{equation}
        \Frame^+_t(\ensemble) \coloneqq \Exp_{U, V \in \ensemble} \, \lvert \tr( \Pi_+ U V^\dagger)\rvert^{2t},
    \end{equation}
    where $\Pi_+$ projects onto the even-parity subspace of $\Hilbert$:
    \begin{equation}
    \label{eq:even_parity_projector}
        \Pi_+  \equiv \frac{1}{2} \left(I + (-1)^F \right).
    \end{equation}
\end{definition}

Any unitary ensemble $\ensemble$ acting on a Hilbert space of dimension $2^{n-1} = \tr(\Pi_+)$ can be extended to a parity-preserving ensemble $\overline{\ensemble}$ in $2^n$ dimensions by having its elements act trivially on the 2-dimensional odd-parity subspace. In this case the parity-restricted frame potential of $\overline{\ensemble}$ is the same as the ordinary one for $\ensemble$:
\begin{equation}
     \Frame^+_t(\overline{\ensemble}) = \Frame_t(\ensemble),
\end{equation}
which confirms that the above definition is reasonable.

The same approach is however not feasible for $\Cliff^{\parity}_{2n}$, as the sub-ensemble acting on the even-parity subspace is not known a priori. Instead, we prove the following:

\begin{theorem}
\label{thm:parity_frame}
    For any subgroup $\Majo^+_{2n} \leq \mathcal{G} \leq \Cliff^{\parity}_{2n}$ of the p-Clifford group, the parity-restricted frame potential $\Frame^+_t(\mathcal{G})$ can be expressed in terms of the subgroup's orthogonal representation $G \leq \Ortho(2n)$:
    \begin{equation}
    \label{eq:parity_frame_fixed_points}
        \Frame^+_t(\mathcal{G}) = \frac{1}{|G|} \sum_{S \in G} \left[\frac{f_+(S) + c_+(S)}{2}  \right]^{t-1},
    \end{equation}
    where $f_+(S)$ is the number of even-parity fixed points of $S$ (satisfying $S v = v$) and $c_+(S)$ is the number of even-parity \emph{complemented points} (satisfying $S v = v^c$).
\end{theorem}

The formal proof is provided in Appendix \ref{sec:parity_frame_proof}, but for now the expression $(f_+(S) + c_+(S)) / 2$ being summed over on the right side of \eqref{eq:parity_frame_fixed_points} has the intuitive interpretation of corresponding to the number of fixed points in the quotient vector space $\F^{2n, +} / \sim^c$ associated to the equivalence relation $v \sim^c v^c = v + j$. This relation makes sense because in the even-parity sector of $\Hilbert$ one has
\begin{equation}
\label{eq:even_parity_Hilbert_identification}
    \majo(v) \ket{\psi} = \majo(v) (-1)^F \ket{\psi} = \pm \majo(v^c) \ket{\psi},
\end{equation}
meaning that $\majo(v)$ and $\majo(v^c)$ act in the same way, up to a sign.  It therefore doesn't matter if $S$ maps $v$ to itself or its complement, which is the reason why one has to also consider $c_+(S)$. The additional factor of 1/2 ensures that each pair $(v, v^c)$ is only counted once, since if $v$ is a fixed/complemented point of $S$, then so is $v^c$.

While the number of even-parity fixed points $f_+(S)$ for a given orthogonal matrix $S \in \Ortho(2n)$ is not known a priori, it is easy to see that the number of even-parity complemented points $c_+(S)$ must be either 0 or equal to $f_+(S)$. One can see this from the fact that the difference of two complemented points is always a fixed point, and therefore a single complemented point suffices to construct $(f_+(S) - 1)$ other such points.

Again using the orbit-stabilizer relation, one might conclude that $\Frame^+_t(\mathcal{G})$ for any subgroup $\Majo^+_{2n} \leq \mathcal{G} \leq \Cliff^{\parity}_{2n}$ is equal to the number of orbits of $G \leq \Ortho(2n)$ on $(\F^{2n, +} / \sim^c)^{\times (t-1)}$. And while this is true, things simplify significantly by closer analyzing how $\Ortho(2n)$ (and its subgroups) act on this quotient space.

\subsection{Closing the Circle}

Comparing the parity-restricted frame potentials of $\Cliff^{\parity}_{2n}$ and $\Cliff_{2n-2} \otimes I_-$ , we found (exactly for $n = 2,3,4$ and approximately for $n \geq 4$) that $\Frame^+_t(\Cliff^{\parity}_{2n})$ (using \eqref{eq:parity_frame_fixed_points}) matches $\Frame^+_t(\Cliff_{2n} \otimes I_-) = \Frame_t(\Cliff_{2n})$ (using \eqref{eq:frame_fixed_points}) precisely. This lead us to formulate an exact proof:

\begin{theorem}
\label{thm:frame_equality}
    The parity-restricted frame potential of $\Cliff^{\parity}_{2n}$ is equal to the ordinary frame potential of $\Cliff_{2n-2}$:
    \begin{equation}
        \Frame^+_t(\Cliff^{\parity}_{2n}) = \Frame_t(\Cliff_{2n-2}) \quad \forall n \geq 2, t \geq 1.
    \end{equation}
\begin{proof}
    Using \eqref{eq:frame_fixed_points} and \eqref{eq:parity_frame_fixed_points}, it is sufficient to show that the effective action of $\Ortho(2n)$ on $\F^{2n, +} / \sim^c$ is equal to ($2^{2n-1}$ copies of) $\Sp(2n-2)$ acting on $\F^{2n-2}$. The proof can be found in Appendix \ref{sec:subgroup_lemma}. This then implies that the distribution of $f(S)$ ($S \in \Sp(2n-2)$) and $(f_+(S) + c_+(S))/2$ ($S \in \Ortho(2n)$) is the same up to an overall multiplicity of $2^{2n-1}$, which then cancels with part of the denominator in \eqref{eq:parity_frame_fixed_points}, since \eqref{eq:symp_group_order} and \eqref{eq:ortho_group_order} imply that
    \begin{equation}
        |\Ortho(2n)| = 2^{2n-1} \cdot |\Sp(2n-2)|.
    \end{equation}
\end{proof}
\end{theorem}
In other words, the subgroup $\Cliff^{\parity}_{2n}$ of parity-preserving Cliffords behaves like $2^{2n-1}$ copies of the ordinary Clifford group $\Cliff_{2n-2}$ when restricted to acting on Majorana strings with even parity and endowed with the identification $\majo(v) \sim \majo(v^c)$ (up to phases) due to \eqref{eq:even_parity_Hilbert_identification}. Each copy acts the same on the even-parity strings and must therefore act differently on a single linearly-independent odd-parity string necessary to generate all of $\Majo_{2n}$. The specific number of copies then arises from there being $2^{2n-1} = |\F^{-, 2n}|$ different ways to map an odd-parity string to another odd-parity string (ignoring phases). Similar statements can probably also be made for subgroups of $\Cliff^{\parity}_{2n}$, but are not of interest to us here.

Overall, while this implies that the p-Clifford group $\Cliff^{\parity}_{2n}$ is indeed a 3-design on the fixed-parity subspaces of $\Hilbert$ for all $n \geq 2$, in terms of practical applications there seems to be no obvious advantage compared to the naive ensemble $\Cliff_{2n-2} \otimes I_\pm$, besides the former being the more natural choice due to not requiring an explicit Hilbert space decomposition. Hence there is also no advantage to simulating these ensembles on a quantum computer employing qubits degrees of freedom. Nevertheless, the recursive relationship between operators with and without parity-preservation seems to provide valuable insights into the structure of the Majorana Clifford group.

\section{Conclusions}
\label{sec:conclusions}

In this paper, we have compiled a comprehensive overview of Majorana strings and the corresponding Clifford operators through the lens of their symplectic representation. Assuming parity superselection, we were able to identify the subgroup of parity-preserving Cliffords, which is represented by binary orthogonal matrices. We also showed how these p-Cliffords can be efficiently generated from a small set of braiding operators (and the parity operator), and how they can be used to prepare any even-parity stabilizer code (assuming access to an ancilla system composed of two Majorana modes). 

Using this framework, a frame potential analysis of the p-Clifford subgroup restricted to a fixed-parity sector of the Hilbert space showed that it is equivalent to the ordinary Clifford group acting on the same subspace. The p-Clifford group interpreted as a random ensemble is thus also a 3-design under the imposed restrictions.

As mentioned in the introduction, one future application of the methods developed in this paper is the study of random tensor network models describing the SYK model, which is envisioned in \cite{bettaque_nora_tensor_network_2024}. But we suspect that other fermionic tensor network models or matchgate circuits (with ties to holography) could also benefit from the toolset established here \cite{jahn_majorana_dimers_holographic_2019, mortier_fermionic_tensor_network_2024, bampounis_matchgate_hierarchy_clifford_2024}.

More generally, our results might prove to be useful for the in-depth analysis of generic classes of systems, such as fermionic quantum codes, (non-)Gaussian states, random circuits and classical shadows \cite{zhao_fermionic_partial_tomography_2021}. In that context it could also prove useful to better understand how the presence of a Hamiltonian can impose restrictions on the possible (Jordan-Wigner) transformations \cite{elman_free_fermions_disguise_2021}, and what relative phase differences can be induced by p-Cliffords when acting on superpositions of Majorana strings and quantum states.

Last but not least, it would be interesting to investigate potential fermionic analogs of the Wigner function or the stabilizer R\'enyi entropy, which are useful tools for characterizing the stabilizerness of bosonic quantum states \cite{gross_hudsons_theorem_finite_2006, leone_stabilizer_renyi_entropy_2022}. But at least for the former there are known difficulties in the qubit case which may be inherited by the fermionic setting.

\section{Acknowledgments}

V.B.\ would like to thank Nadie LiTenn and Shiyong Guo for some insightful discussions and comments. We acknowledge support from  the AFOSR under grant number FA9550-19-1-0360 (V.B. and B.G.S.).

\bibliographystyle{quantum}
\bibliography{refs.bib}

\appendix

\section{Canonical Jordan-Wigner Transformation}
\label{sec:canonical_jordan_wigner}

Based on the $\F^{2n}$ representation $W$ of the Jordan-Wigner transformation \eqref{eq:jordan_wigner_matrix}, we provide a specific canonical choice in terms of the actual Hilbert space operators. This was done by reading off the columns of $W$ and plugging them into definitions \eqref{eq:majo_string} and \eqref{eq:pauli_string} respectively, thereby implicitly fixing the signs of the mapping. For the latter we also employed the fact that $i X Z = Y$. Using the short-hand tensor product notation
\begin{align}
\begin{split}
    P_k &\equiv I^{\otimes (k-1)} \otimes P \otimes I^{\otimes (n-k)}
\end{split}
\end{align}
for $P = X, Y, Z$ and $k = 1,\ldots,n$, the resulting transformations are then as follows:

~\\
\textbf{Majorana to Pauli:}
\begin{align}
\begin{split}
    \chi_{2k-1} &= Y_1 Y_2 \cdots Y_{k-1} \, X_k, \\
    \chi_{2k} &= Y_1 Y_2 \cdots Y_{k-1} \, Z_k.
\end{split}
\end{align}
\\
\textbf{Pauli to Majorana:}
\begin{align}
\begin{split}
    X_k &=  (i)^{(k-1) \text{ mod } 2} \cdot \chi_1 \chi_2 \cdots \chi_{2k-2} \, \chi_{2k-1}, \\
    Z_k &= (i)^{(k-1) \text{ mod } 2} \cdot \chi_1 \chi_2 \cdots \chi_{2k-2} \, \chi_{2k}.
\end{split}
\end{align}

\section{Majorana Algebras}
\label{sec:majorana_algebras}

Any element $O$ of a Majorana algebra acting on a $2^n$-dimensional Hilbert space $\Hilbert$ can be expressed as a linear combination of Majorana strings, such that
\begin{equation}
\label{eq:algebra_expansion}
    O \equiv \sum_{v \in \F^{2n}} c_v \cdot \majo(v), \quad c_v \in \mathbb{C}.
\end{equation}
In fact, it is straightforward to show that a Majorana algebra composed of $2n$ modes encompasses all complex-valued $2^n \times 2^n$ matrices \cite{benatti_entanglement_algebraic_quantum_2016}. Using  \eqref{eq:majo_trace} and the fact that Hermitean strings square to the identity, the coefficients $c_v$ of the expansion can be determined using:
\begin{equation}
    c_v = \frac{1}{2^n} \tr(O \majo(v)).
\end{equation}

A Majorana subalgebra $\mathcal{A}$ is any algebra spanned by Majorana strings which only contain an (ordered) even subset $A \subseteq I$ of all modes $I = \{1, 2, \ldots, 2n\}$ (here represented by their respective indices). These strings are therefore of the form 
\begin{equation}
\label{eq:naive_subalgebra_strings}
    \majo(\Pi(v_A \oplus 0_{\Bar{A}})), \quad v_A \in \F^{2n_A},
\end{equation}
where $n_A = |A|/2$, and $\Pi$ is some fixed permutation matrix. Note however that we run into issues with regard to the Hilbert space representation of the subalgebra. The strings as depicted in \eqref{eq:naive_subalgebra_strings} are still endowed with the ordinary $2^n \times 2^n$-representation \eqref{eq:majo_trace}, even though the algebra itself is isomorphic to the $2^{n_A} \times 2^{n_A}$ algebra of complex matrices. Hence this representation is reducible, which is especially problematic when considering properties like entropy that intrinsically depend on the Hilbert space dimension. To compensate, one has introduce a new irreducible representation $\majo_A(v_A)$ which satisfies \eqref{eq:majo_string} and \eqref{eq:majo_trace} with $n \rightarrow n_A$. The reduced operator $O_A$ corresponding to the chosen subalgebra is then given by
\begin{align}
\begin{split}
    O_A &= \sum_{v_A} c_{v_A} \cdot \majo_A(v_A), \\
    c_{v_A} &= \frac{1}{2^n} \tr(O \majo(\Pi(v_A \oplus 0_{\Bar{A}}))).
\end{split}
\end{align}

Conversely, one can define the algebraic tensor product $\otimes_f$ of two algebras linearly in terms of its action on individual strings:
\begin{equation}
    \majo_A(v_A) \otimes_f \majo_B(v_B) \equiv \majo_{AB}(v_A \oplus v_B),
\end{equation}
where $\majo_A, \majo_B$ and $\majo_{AB}$ represent the irreducible representations of their respective algebras. Note however that this operation is only really well-defined for even-parity strings, as otherwise the two possible orders in which to take the product will differ by an overall sign.

\section{Examples}

\subsection{Fermionic Permutation Operators}
\label{sec:fermionic_swap}

Weight-2 braiding operators \eqref{eq:braid_operators} act as simple permutation operators. To see this, let for example:
\begin{equation}
    \Pi_{ij} \coloneqq \exp\left(-\frac{\pi}{4} \chi_i \chi_j\right) \equiv \braid(i \chi_i \chi_j) \quad (i \neq j).
\end{equation}
Then from \eqref{eq:braiding_action} (or direct computation) it follows that the non-trivial action of $\Pi_{ij}$ on single Majorana modes is fully characterized by
\begin{equation}
\label{eq:fermionic_permutation_action}
    \Pi_{ij} \chi_i \Pi_{ij}^\dagger = \chi_j, \quad  \Pi_{ij} \chi_j \Pi_{ij}^\dagger = -\chi_i.
\end{equation}
In that sense, $\Pi_{ij}$ is the equivalent of the SWAP operator for qubits, and occurs in the context of e.g.\ computing Rényi entropies using the replica trick \cite{bentsen_approximate_quantum_codes_2024}, or the ground state of a simple Kitaev chain \cite{kitaev_unpaired_majorana_fermions_2001, jahn_majorana_dimers_holographic_2019}. Products of these fermionic swap operators thus generate a subgroup of p-Cliffords corresponding to the symmetric group $S_{2n} \leq O(2n)$, i.e.\ the group of permutations. These permutations trivially preserve the operator weight of all Majorana strings.

The Cliffords corresponding to elements $S \in S_{2n}$ are a special class of \emph{fermionic Gaussian unitaries}, which are defined as acting according to \cite{bloch_canonical_form_antisymmetric_1962}
\begin{equation}
    U \chi_i \, U^\dagger = \sum_{j = 1}^{2n} O_{ij} \chi_j,
\end{equation}
where in general $O$ can be any \emph{real} $2n \times 2n$ orthogonal matrix, but here must be equivalent to $S$, up to potential factors of $-1$ in some entries due to \eqref{eq:fermionic_permutation_action}.

Furthermore, one can also ask which subgroup of the Gaussian Cliffords\footnote{It is easy to see that non-Gaussian Cliffords can't preserve $\hat{N}$.} preserves the fermionic particle number operator
\begin{align}
\begin{split}
\label{eq:number_operator}
    \hat{N} &\coloneqq \sum_{i=1}^{n} c_i^\dagger c_i \\
    &= \frac{1}{2} \left( n \, I + i \sum_{i=1}^{n} \chi_{2i-1} \chi_{2i} \right),
\end{split}
\end{align}
defined in terms of the canonical creation and annihilation operators $c_i \equiv (\chi_{2i-1} + i \chi_{2i}) / 2$ and $c_i^\dagger \equiv (\chi_{2i-1} - i \chi_{2i}) / 2$. Since in this case the only possible non-trivial thing the Cliffords can do is permute the terms of the sum, the symplectic action must be $S_n \otimes I_2 \leq S_{2n}$. Potential sign differences such as in \eqref{eq:fermionic_permutation_action} can then be avoided due to the modes in \eqref{eq:number_operator} always occurring in pairs.

\subsection{The Shortest Fermion Code}
\label{sec:shortest_fermion_code}

The shortest fermion code is a $[[6, 1, 3]]_f$ stabilizer code, meaning it utilizes 12 Majorana modes to encode one logical qubit, with the ability to correct one error and detect up to two \cite{vijay_quantum_error_correction_2017}. One possible stabilizer basis for it is given by
\begin{align}
\begin{split}
    \sigma_1 &= \chi_1 \chi_2 \chi_3 \chi_4, \\
    \sigma_2 &= \chi_3 \chi_4 \chi_5 \chi_6, \\
    \sigma_3 &= \chi_7 \chi_8 \chi_9 \chi_{10}, \\
    \sigma_4 &= \chi_9 \chi_{10} \chi_{11} \chi_{12}, \\
    \sigma_5 &= i \chi_2 \chi_4 \chi_6 \chi_8 \chi_{10} \chi_{12}. \\
\end{split}
\end{align}
In the symplectic representation, this corresponds to the following binary basis matrix (and $v = 0$):
\begin{equation}
\label{eq:shortest_fermion_code_matrix}
    M = \begin{pmatrix}
        1 & 1 & 1 & 1 & 0 & 0 & 0 & 0 & 0 & 0 & 0 & 0 \\
        0 & 0 & 1 & 1 & 1 & 1 & 0 & 0 & 0 & 0 & 0 & 0 \\
        0 & 0 & 0 & 0 & 0 & 0 & 1 & 1 & 1 & 1 & 0 & 0 \\
        0 & 0 & 0 & 0 & 0 & 0 & 0 & 0 & 1 & 1 & 1 & 1 \\
        0 & 1 & 0 & 1 & 0 & 1 & 0 & 1 & 0 & 1 & 0 & 1 \\
    \end{pmatrix}.
\end{equation}
The space of logical operators is spanned by the (unitary) Majorana strings
\begin{align}
\begin{split}
    \ell_1 &= \chi_1 \chi_3 \chi_5, \\
    \ell_2 &= \chi_2 \chi_4 \chi_6 \chi_7 \chi_8 \chi_9 \chi_{10} \chi_{11} \chi_{12},
\end{split}
\end{align}
which when multiplied together result in the total parity operator
\begin{equation}
    -\ell_1 \ell_2 = (-1)^F.
\end{equation}
$(-1)^F$ not being an element of the stabilizer implies that any two independent logical states must have opposite parity, which can be flipped by applying $\ell_1$ and $\ell_2$.

Each weight-1 error $E \sim \majo(e)$ corresponding to quasiparticle poisoning can be checked against by a unique combination of stabilizers basis elements. Those combinations can be neatly read off of \eqref{eq:shortest_fermion_code_matrix} by checking which row vectors $m$ have a non-zero entry in a given column, as for those one has $\inner{m}{e} = m^T e = 1$, meaning that the corresponding operators anticommute with $E$.

Since $(-1)^F$ is not an element of the shortest fermion code, one can use Algorithm \eqref{alg:clifford_from_stabilizer} directly to get an embedding p-Clifford, without having to introduce ancillary Majorana modes. The orthogonal matrix corresponding to this Clifford is given by
\begin{equation}
    S = \begin{pmatrix}
        1 & 0 & 0 & 0 & 0 & 0 & 1 & 1 & 1 & 1 & 0 & 0 \\
        1 & 0 & 1 & 1 & 0 & 0 & 0 & 0 & 0 & 1 & 1 & 0 \\
        0 & 1 & 0 & 1 & 0 & 0 & 1 & 1 & 0 & 0 & 1 & 0 \\
        1 & 0 & 1 & 0 & 0 & 0 & 0 & 0 & 1 & 0 & 0 & 0 \\
        0 & 0 & 1 & 0 & 0 & 0 & 1 & 1 & 1 & 1 & 0 & 0 \\
        1 & 1 & 1 & 0 & 0 & 0 & 0 & 0 & 0 & 1 & 1 & 0 \\
        0 & 0 & 0 & 0 & 1 & 0 & 0 & 0 & 0 & 0 & 0 & 0 \\
        1 & 1 & 1 & 1 & 0 & 1 & 1 & 1 & 0 & 1 & 0 & 1 \\
        1 & 1 & 1 & 1 & 0 & 1 & 0 & 1 & 0 & 0 & 1 & 0 \\
        0 & 0 & 0 & 0 & 0 & 1 & 1 & 0 & 0 & 1 & 1 & 1 \\
        1 & 1 & 1 & 1 & 0 & 0 & 0 & 1 & 0 & 0 & 1 & 1 \\
        0 & 0 & 0 & 0 & 0 & 0 & 0 & 1 & 0 & 1 & 1 & 0 \\
    \end{pmatrix}.
\end{equation}

\section{Further Proofs}

\subsection{Theorem \ref{thm:clifford_symplectic}}
\label{sec:stone_neumann_proof}

\begin{proof} 
We prove each of the three statements as follows:
\begin{enumerate}
    \item Using the fact that $\cliff$ is unitary, one can rewrite \eqref{eq:majorana_commutation} by applying $\cliff (\cdot) \cliff^\dagger$ to both sides and inserting $I = \cliff^\dagger \cliff$:
    \begin{align}
    \begin{split}
        & (\cliff \majo(v) \cliff^\dagger) (\cliff \majo(v') \cliff^\dagger) \\ ={}& (-1)^{\inner{v}{v'}} (\cliff \majo(v') \cliff^\dagger) (\cliff \majo(v) \cliff^\dagger)
    \end{split}
    \end{align}
    Evaluating both sides using \eqref{eq:clifford_group_def} and dividing by $c(v) \, c(v') \neq 0$ then leads to
    \begin{align}
    \begin{split}
         & \majo(S(v)) \majo(S(v')) \\
         ={}& (-1)^{\inner{v}{v'}} \majo(S(v')) \majo(S(v)),
    \end{split}
    \end{align}
    On the other hand, we can also directly rewrite \eqref{eq:majorana_commutation} with the substitution $v \rightarrow S(v)$:
    \begin{align}
    \begin{split}
         & \majo(S(v)) \majo(S(v')) \\
         ={}& (-1)^{\inner{S(v)}{S(v')}} \majo(S(v')) \majo(S(v)).
    \end{split}
    \end{align}
    But since both expressions have to be equivalent, we get the following requirement:
    \begin{equation}
        \inner{S(v)}{S(v')} \stackrel{!}{=} \inner{v}{v'} \quad \forall v, v' \in \F^{2n}.
    \end{equation}
    Linearity of $S$ is then a consequence of $\inner{\cdot}{\cdot}$ defining a linear inner product:
    \begin{align}
    \begin{split}
        & \inner{S(v + v')}{S(v'')} \\ 
        ={}& \inner{v + v'}{v''} \\
        ={}& \inner{v}{v''} + \inner{v'}{v''} \\
        ={}& \inner{S(v)}{S(v'')} + \inner{S(v')}{S(v'')} \\
        ={}& \inner{S(v) + S(v')}{S(v'')} \quad \forall v, v', v'' \in \F^{2n}.
    \end{split}
    \end{align}
    With $S$ being linear it then immediately follows from Definition \ref{def:symplectic} that it must also be symplectic.

    The converse direction follows from the fact that both $\majo(v)$ and $\majo(Sv)$ define irreducible representations of $\F^{2n}$ because of Lemma \ref{thm:majo_irreducible} and $S$ being linear:
    \begin{equation}
        \frac{1}{|\F^{2n}|} \sum_{Sv \in \F^{2n}} |\tr(\majo(S v))|^2 = 1.
    \end{equation}
    Both representations are therefore equivalent and hence (up to an overall phase) connected by a unitary transformation $\cliff$ that induces $S$:
    \begin{align}
    \begin{split}
        1 &= \frac{1}{|\F^{2n}|} \sum_{Sv \in \F^{2n}} |\tr(\majo(S v))|^2 \\
        &= \frac{1}{|\F^{2n}|} \sum_{v \in \F^{2n}} |\tr(\cliff \majo(v) \cliff^\dagger)|^2 \\
        &= \frac{1}{|\F^{2n}|} \sum_{v \in \F^{2n}} |\tr( \majo(v) \cliff^\dagger \cliff)|^2 \\
        &= \frac{1}{|\F^{2n}|} \sum_{v \in \F^{2n}} |\tr( \majo(v))|^2.
    \end{split}
    \end{align}

    \item Assuming $\cliff(S)$ satisfies \eqref{eq:clifford_group_def}, then $\majo(a) \cliff(S)$ with $a \in \F^{2n}$ does so as well:
    \begin{align}
    \begin{split}
    \label{eq:cliff_majo_action}
        & (\majo(a) \cliff(S)) \majo(v) (\majo(a) \cliff(S))^\dagger \\
        ={}& \majo(a) \cliff(S) \majo(v) \cliff^\dagger(S) \majo(a) \\
        ={}& c(v) \cdot \majo(a) \majo(S v) \majo(a) \\
        ={}& (-1)^{\inner{Sv}{a}} c(v) \cdot \majo(S v) \\
        \equiv{}& c'(v) \cdot \majo(S v),
    \end{split}
    \end{align}
    where we used \eqref{eq:majorana_commutation} and the fact that $\majo(a)^2 = I$.

    Conversely, let $\cliff(S)$ and $\cliff'(S)$ be two Cliffords with the same symplectic action. Then dividing both cases of \eqref{eq:cliff_coeff_composition} yields
    \begin{equation}
        \left( \frac{c(v)}{c'(v)} \right) \cdot \left( \frac{c(v')}{c'(v')} \right) = \frac{c(v + v')}{c'(v + v')}
    \end{equation}
    for all $v, v' \in \F^{2n}$. This together with $c(v) = \pm 1$ (and the symplectic form being non-degenerate) implies that -- without loss of generality -- $c(v) / c'(v)$ can be written as 
    \begin{equation}
        \frac{c(v)}{c'(v)} \equiv (-1)^{\inner{v}{a'}} = (-1)^{\inner{S v}{a}},
    \end{equation}
    for some $a' \equiv S^{-1} a \in \F^{2n}$ due to \eqref{eq:symp_group_def}. This coincides with the form of the phase induced by an additional Majorana string $\majo(a)$, as shown in \eqref{eq:cliff_majo_action}.

    \item $\Cliff_{2n} / \sim$ forms a representation of $\Sp(2n)$, because:
    \begin{align}
    \begin{split}
        & \cliff(S) \cliff(T) \majo(v) \cliff^\dagger(T) \cliff^\dagger(S) \\
        \propto{}& \cliff(S) \majo(T v) \cliff^\dagger(S) \\
        \propto{}& \majo(ST v) \\
        \propto{}& \cliff(ST) \majo(v) \cliff^\dagger(ST) \quad \forall v \in \F^{2n},
    \end{split}
    \end{align}
    which implies that $\cliff(S) \cliff(T) \sim \cliff(ST)$ and therefore $[\cliff(S)] [\cliff(T)] = [\cliff(ST)]$.
\end{enumerate}
\end{proof}

\subsection{Theorem \ref{thm:parity_frame}}
\label{sec:parity_frame_proof}

\begin{proof}
    Because $\mathcal{G} \leq \Cliff^{\parity}_{2n}$ forms a group and therefore satisfies $U V^\dagger \in \mathcal{G}$ for all $U,V \in \mathcal{G}$, the frame potential can be simplified to be 
    \begin{align}
    \begin{split}
    \label{eq:parity_frame_potential_group}
        \Frame^+_t(\mathcal{G}) &= \Exp_{U \in \mathcal{G}} \, \lvert \tr( \Pi_+ U)\rvert^{2t} \\
        &= \frac{1}{\lvert\F^{2n, +}\rvert} \frac{1}{|G|} \sum_{S \in G} \sum_{a \in \F^{2n, +}} \lvert \tr[\Pi_+ \cliff(S) \majo(a)] \rvert^{2t}.
    \end{split}
    \end{align}
    Next we first show that $\lvert \tr(\Pi_+ U) \rvert^2$ with $U = \majo(a) \cliff(S) \in \mathcal{G}$ is either zero or equal to $(f_+(S) + c_+(S)) / 2$. For that we evaluate the expression as a sum over the Bell basis
    \begin{equation}
        \ket{\Phi_n(v)} \equiv (\majo(v) \otimes I) \ket{\Phi_n^+} \in \Hilbert \otimes \Hilbert,
    \end{equation}
    where $\ket{\Phi_n^+} \equiv \ket{\Phi^+}^{\otimes n}$ denotes $n$ copies of canonical Bell state. That these states indeed form a complete basis of $\Hilbert \otimes \Hilbert$ follows from:
    \begin{align}
    \begin{split}
        \braket{\Phi_n(v) | \Phi_n(v')} &= \braket{\Phi_n^+ | \majo(v) \majo(v') \otimes I | \Phi_n^+} \\
        &= \frac{1}{2^n} \tr(\majo(v) \majo(v')) \\
        &= \delta_{v, v'},
    \end{split}
    \end{align}
    (using \eqref{eq:majo_trace} and a well-known property of the canonical Bell state) and the fact that $\dim(\Hilbert \otimes \Hilbert) = 2^{2n} = |\F^{2n}|$. With that we can evaluate the aforementioned trace:
    \begin{align}
    \begin{split}
    \label{eq:trace_squared_parity_1}
        & \lvert \tr(\Pi_+ U) \rvert^2 \\
        ={}& \tr(\Pi_+ U \otimes \Pi_+ U^\dagger) \\
        ={}& \sum_{v \in \F^{2n}} \bra{\Phi_n(v)} \Pi_+ U \otimes \Pi_+ U^\dagger\ket{\Phi_n(v)} \\ 
        ={}& \sum_{v \in \F^{2n}} \bra{\Phi_n^+} \majo(v) \Pi_+ U \majo(v) \otimes \Pi_+ U^\dagger \ket{\Phi_n^+} \\
        \stackrel{\text{(a)}}{=} {}& \sum_{v \in \F^{2n}} \bra{\Phi_n^+} \majo(v) \Pi_+ U \majo(v) U^\dagger \Pi_+ \otimes I \ket{\Phi_n^+} \\
        ={}& \sum_{v \in \F^{2n}} c(v) \,\bra{\Phi_n^+} \majo(v) \Pi_+ \majo(S v) \Pi_+ \otimes I \ket{\Phi_n^+} \\
        \stackrel{\text{(b)}}{=} {}& \sum_{v \in \F^{2n, +}} c(v) \,\bra{\Phi_n^+} \majo(v) \majo(S v) \Pi_+ \otimes I \ket{\Phi_n^+} \\
        ={}& \sum_{v \in \F^{2n, +}} \frac{c(v)}{2^n} \, \tr [ \majo(v) \majo(S v) \Pi_+ ] \\
        \stackrel{\text{(c)}}{=} {}& \sum_{v \in \F^{2n, +}} \frac{c(v)}{2^{n+1}} \, \tr[\majo(v) \majo(S v) + \majo(v) \majo(S v) \majo(j)] \\
        \stackrel{\text{(d)}}{=} {}& \sum_{v \in \F^{2n, +}} \frac{c(v)}{2} \, ( \delta_{Sv, v} + \zeta(j, v) \, \delta_{Sv, v^c} ).
    \end{split}
    \end{align}
    The steps labeled in the derivation are achieved in the following ways:
    \begin{enumerate}[(a)]
        \item The equivalence is straightforward to prove by expressing both inner products using circuit notation (also called Penrose graphical notation), akin to how it is done for quantum teleportation. 
        \item $\Pi_+ \majo(Sv) \Pi_+$ either vanishes if $Sv$ has odd parity, or is equal to $\majo(Sv) \Pi_+$ if $Sv$ has even parity. And since $S$ preserves parity, this means that the sum over $v \in \F^{2n}$ reduces to the sum over $v \in \F^{2n, +}$.

        \item We insert the definition \eqref{eq:even_parity_projector} of the projector $\Pi_+$ with the identification $(-1)^F \sim \majo(j).$

        \item We fuse $\majo(v)$ and $\majo(j)$ using trace cyclicity and \eqref{eq:majorana_composition}, and then apply \eqref{eq:majo_trace} to both terms.
    \end{enumerate}
    We can now analyze the structure of $c(v)$ for the resulting cases of $Sv = v$ and $Sv = v^c$:
    \begin{itemize}
        \item In the case of $S v = v$, it follows directly from \eqref{eq:cliff_coeff_composition} that the coefficients of any two (even-parity) fixed points $v, v'$ satisfy $c(v) \cdot c(v') = c(v + v')$. This together with \eqref{eq:clifford_phase_pm} and \eqref{eq:cliff_majo_action} implies that we can identify $c(v) \equiv (-1)^{\inner{Sv}{a}} = (-1)^{a^T v}$, up to an overall sign that might arise to make the relevant terms in \eqref{eq:trace_squared_parity_1} positive, but which can be ignored without loss of generality.

        \item In the case of $S v = v^c$, \eqref{eq:cliff_coeff_composition} can be rewritten as
        \begin{equation}
        \label{eq:cliff_coeff_compo_compl}
            c(v) \cdot c(v') = \frac{\zeta(v, v')}{\zeta(v^c, (v')^c)} \cdot c(v + v'),
        \end{equation}
        for any two (even-parity) complemented points $v, v'$.
        But since the sum of two such points is always a fixed point, we can identify $c(v + v') = (-1)^{a^T (v + v')}$ as before. Additionally, by evaluating $\majo(j) \majo(v) \majo(v') \majo(j)$ in two different orders one can arrive at
        \begin{equation}
            \zeta(v, v') = \zeta(j, v) \cdot \zeta(j, v') \cdot \zeta(v^c, (v')^c).
        \end{equation}
        Plugging this all into \eqref{eq:cliff_coeff_compo_compl} and using $\zeta(j, v) = \pm 1$ (since $\inner{j}{v} = 0$), results in
        \begin{equation}
            \left( c(v) \, \zeta(j, v) \right) \cdot \left( c(v') \, \zeta(j, v') \right) = (-1)^{a^T (v + v')},
        \end{equation}
        which therefore implies $c(v) \, \zeta(j, v) = (-1)^{a^T v}$.
    \end{itemize}

    Applying these results to \eqref{eq:trace_squared_parity_1} then leaves us with
    \begin{align}
    \begin{split}
    \label{eq:trace_squared_parity_2}
        \lvert \tr(\Pi_+ U) \rvert^2 &= \sum_{v \in \F^{2n, +}} \frac{(-1)^{a^T v}}{2} ( \delta_{Sv, v} + \delta_{Sv, v^c} ) \\ 
        &= \frac{1}{2} \left(f_+(S) + c_+(S)\right) \quad (\text{or } 0),
    \end{split}
    \end{align}
    where the second line is a direct consequence of the discrete Fourier transform identity \cite{gross_hudsons_theorem_finite_2006}
    \begin{equation}
    \label{eq:fourier_orthogonal}
        \sum_{v \in M} (-1)^{a^T v} = |M| \, \delta_{a \in M^{\perp}},
    \end{equation}
    which holds for any subspace $M \subseteq \F^{2n}$ and the corresponding orthogonal space.
    \begin{equation}
        M^\perp \coloneqq \{ v \in \F^{2n} \, | \, v^T v' = 0 \quad  \forall v' \in M \}.
    \end{equation}
   Equation \eqref{eq:fourier_orthogonal} also implies that
   \begin{align}
   \begin{split}
   \label{eq:sum_trace_squared_parity}
        & \sum_{a \in \F^{2n, +}} \lvert \tr[\Pi_+ \cliff(S) \majo(a)] \rvert^2 \\
        ={}& \sum_{v \in \F^{2n, +}} \sum_{a \in \F^{2n, +}} \frac{(-1)^{a^T v}}{2} ( \delta_{Sv, v} + \delta_{Sv, v^c} ) \\
        ={}& \lvert \F^{2n, +} \rvert \sum_{v \in \F^{2n, +}}  \frac{1}{2} ( \delta_{Sv, v} + \delta_{Sv, v^c} ) \, \delta_{v \in (\F^{2n, +})^\perp} \\
        ={}& \lvert \F^{2n, +} \rvert,
    \end{split}
    \end{align}
    since $(\F^{2n, +})^\perp \cap \F^{2n, +} = \{0, j\}$, and both elements are fixed points.
    Hence the sum in the beginning of \eqref{eq:sum_trace_squared_parity} must have $2 \cdot \lvert \F^{2n, +} \rvert / (f_+(S) + c_+(S))$ non-zero terms, each of which is equal to \eqref{eq:trace_squared_parity_2}. It follows that
    \begin{align}
    \begin{split}
         & \sum_{a \in \F^{2n, +}} \lvert \tr[\Pi_+ \cliff(S) \majo(a)] \rvert^{2t} \\
         ={}& \frac{2 \, \lvert \F^{2n, +} \rvert}{(f_+(S) + c_+(S))} \cdot \left[\frac{f_+(S) + c_+(S)}{2} \right]^t \\
         ={}& \lvert\F^{2n, +}\rvert \cdot \left[\frac{f_+(S) + c_+(S)}{2} \right]^{t-1},
    \end{split}
    \end{align}
    which when plugged into \eqref{eq:parity_frame_potential_group} yields
    \begin{equation}
        \Frame^+_t(\mathcal{G}) = \ \frac{1}{|G|} \sum_{S \in G}  \left[\frac{f_+(S) + c_+(S)}{2} \right]^{t-1},
    \end{equation}
    thus completing the proof.
\end{proof}

\subsection{Lemma for Theorem \ref{thm:frame_equality}}
\label{sec:subgroup_lemma}

\begin{lemma}
    The orthogonal group $\Ortho(2n)$ restricted to $\F^{2n, +} / \sim^c$ (where $v \sim^c v^c = v + j$) is equal to ($2^{2n-1}$ copies of) $\Sp(2n-2)$, and therefore $\Sp(2n-2) \leq \Ortho(2n)$.
\begin{proof}
    Without loss of generality, we start with the following basis of $\F^{2n, +}$ as a $(2n-1) \times 2n$ matrix, with its elements as row vectors:
    \begin{equation}
        B = \begin{pmatrix}
            1 & 0 & 1 & 0 & 0 & \cdots & 0 & 0 & 0 & 0 \\
            1 & 1 & 0 & 0 & 0 & \cdots & 0 & 0 & 0 & 0 \\
            1 & 1 & 1 & 0 & 1 & \cdots & 0 & 0 & 0 & 0 \\
            1 & 1 & 1 & 1 & 0 & \cdots & 0 & 0 & 0 & 0 \\
            \vdots & \vdots & \vdots & \vdots & \vdots & \ddots & \vdots & \vdots & \vdots & \vdots \\
            1 & 1 & 1 & 1 & 1 & \cdots & 1 & 0 & 1 & 0 \\
            1 & 1 & 1 & 1 & 1 & \cdots & 1 & 1 & 0 & 0 \\
            \hline 1 & 1 & 1 & 1 & 1 & \cdots & 1 & 1 & 1 & 1
        \end{pmatrix}.
    \end{equation}
    Going to $\F^{2n, +} / \sim^c$ is then achieved by removing the last line of $B$ (corresponding to $j$) and identifying all other lines with their respective complement\footnote{Identifying each basis element with its complement is only really necessary to ensure the basis is still complete, it does not affect anything else.}. We will refer to this new basis as $\widetilde{B}$, and any element of $\F^{2n, +} / \sim^c$ can be expressed as $\widetilde{B} ^T v$ with $v \in \F^{2n -2}$. The canonical inner product of any two elements then is
    \begin{equation}
        \left(\widetilde{B}^T v \right)^T \left(\widetilde{B}^T v'\right) = v^T \left(\widetilde{B} \widetilde{B}^T\right) v' = v^T \symppauli v',
    \end{equation}
    where $\symppauli$ is the symplectic form in the Pauli basis \eqref{eq:symp_matrix_pauli}. For any orthogonal matrix $S \in \Ortho(2n)$ there therefore must be a symplectic matrix $\widetilde{S} \in \Sp(2n-2)$ such that
    \begin{equation}
        \widetilde{B} S^T = \widetilde{S}^T \widetilde{B}.
    \end{equation}
    Conversely, for any $\widetilde{S} \in \Sp(2n-2)$ there are $2^{2n-1}$ orthogonal matrices $S \in \Ortho(2n)$ with that effective action on $\F^{2n, +} / \sim^c$, because they can only differ by how they map the one-dimensional odd-parity subspace to one of the $2^{2n-1}$ possible options.
\end{proof}
\end{lemma}

\section{Sampling p-Cliffords}
\label{sec:sampling}

In this appendix we provide an efficient sampling algorithm for elements of the p-Clifford group $\Cliff^{\parity}_{2n}$ in terms of their orthogonal representation $S \in \Ortho(2n)$. It is inspired by the more general algorithm for sampling arbitrary symplectics provided in \cite{koenig_how_efficiently_select_2014}.

Since orthogonal matrices exist for any even or odd dimension $N \geq 1$, the algorithm can be defined recursively to sample all of them. It uses the fact that any two vectors $v, w \in \F^{N} \setminus \{j, 0\}$ of the same parity are connected by either one or two Householder reflections $h_1, h_2 \in \Ortho(N)$, such that
\begin{equation}
    h_2 h_1 v = w,
\end{equation}
where $h_2$ is potentially just the identity.

\begin{figure}[hbt]
\begin{algorithm}[H]
\caption{HOUSEHOLDER($v, w$)}\label{alg:finding_householders}
\begin{algorithmic}
    \Require $v, w \in \F^N, \, \parity(v) = \parity(w), \, v \neq j \neq w$
    \Ensure $a, b \in \F^{N, +}, \, h_b h_a v = w$

    \If{$v^T w = 1 - \parity(v)$ or $v = w$}
        \State $a \gets w - v$
        \State $b \gets 0$
        \State \Return
    \EndIf
    
    \State $a \gets 0$    

    \For{$i = 1,\ldots,N$}
        \If{$v_i = 0$ and $w_i = 0$}
            \For{$i' = 1,\ldots,N$}
            \If{$v_{i'} = 1$ and $w_{i'} = 1$}
                \State $a_i \gets 1, \, a_{i'} \gets 1$
                \State $b \gets w - v - a$
                \State \Return
        \EndIf
    \EndFor
     \State \Break
        \EndIf
    \EndFor

    \For{$i = 1,\ldots,N$} 
        \If{$v_i = 1$ and $w_i = 0$}
            \State $a_i \gets 1$
            \State \Break 
        \EndIf
    \EndFor
     \For{$i = 1,\ldots,N$}
        \If{$v_i = 0$ and $w_i = 1$}
            \State $a_i \gets 1$
            \State \Break 
        \EndIf
    \EndFor
    \State $b \gets w - v - a$
    \State \Return
\end{algorithmic}
\end{algorithm}
\end{figure}

This algorithm has $\mathcal{O}(N)$ runtime and returns the even-parity vectors defining the Householder reflections connecting $v$ and $w$. It acts as follows:
\begin{enumerate}
    \item We first check if $v^T w = 1 - \parity(v)$ (or $v = w$), because then $v^T(w - v) = 1$ (or $v^T(w - v) = 0$ if $v = w$ since nothing has to be done), and hence the Householder $h_{w - v}$ satisfies the desired property.

    \item Next we check if $v$ and $w$ have a common index $i$ for which $v_i = 0 = w_i$. If such an index exists, we also check if $v$ and $w$ share an index $i'$ for which $v_{i'} = 1 = w_{i'}$. If both indices exist, then $a$ with $a_i = 1 = a_{i'}$ as the only non-zero elements satisfies $a^T v = 1 = a^T w$ and thus together with $b = w - v - a$ provides a valid pair or reflections. 

    \item We can now assume that $v$ is not fully contained in $w$, or vice versa. If that were the case, they would have a common zero since neither of them can be equal to $j$. Therefore the only other option is that $v_i = 1, w_i = 0$ for (at least) one index $i$, and $v_{i'} = 0, w_{i'} = 1$ for another index $i'$. The vector $a$ that only has ones at $i$ and $i'$ -- and zeroes everywhere else -- therefore describes a valid reflection. The other one is again described by $b = w - v - a$.  
\end{enumerate}

The actual routine for sampling orthogonal matrices is now straightforward. It is $\mathcal{O}(N^3)$ because it has $N$ recursive steps, during each of which it takes $\mathcal{O}(N)$ to find the Householder reflection that maps $(1,0,\ldots,0)^T \in \F^N$ to the (odd-parity) binary expansion corresponding to the leading $N-1$ bits of $1 \leq i \leq |\Ortho(N)|$. Applying these reflections to said basis vector (and the other $\mathcal{O}(N)$ ones from the previous recursion steps) then takes $\mathcal{O}(N)$ steps each, since doing so only involves computing inner products and adding vectors. Overall it therefore requires $\mathcal{O}(N^2)$ time to apply the Householders to all columns. Uniqueness of this mapping follows from the ability to perform all these steps in reverse, i.e.\ by reading the binary representation of the corresponding index $i$ off the the leading column vectors at each inverse recursion step.

\begin{figure}[hbt!]
\begin{algorithm}[H]
\caption{ORTHOGONAL($N, i$)}\label{alg:orthogonal_sampling}
\begin{algorithmic}
    \Require $1 \leq i \leq |\Ortho(N)|$
    \Ensure $g_i \in \Ortho(N), \quad g_i \neq g_{i'} \Leftrightarrow i \neq i' $

    \If{$N = 1$}
        \State $g_i \gets 1$
        \State \Return
    \EndIf
    \State $p \gets 2^{N - 1}$ \Comment{Number of odd-parity vectors of length $N$}
    \State $i_p \gets i - 1 \mod p$
    \State $i_r \gets (i - 1) / p + 1$ \Comment{Index for next recursive step.}
    \State $b_i \gets \text{binary representation of } i_p \text{ as vector in } \F^{N-1}$ 
    \State $e \gets (1, 0, \ldots, 0)^T \in \F^N$ 
    \State $f \gets (\parity(b_i) + 1) \oplus b_i \in \F^N$
    \State $a, b \gets \text{HOUSEHOLDER}(e, f)$
    \State $g_i \gets h_b h_a (1 \oplus \text{ORTHOGONAL}(N-1, i_r))$
    \State \Return
\end{algorithmic}
\end{algorithm}
\end{figure}

\end{document}